\documentclass[twocolumn,preprintnumbers,amsmath,amssymb,floatfix]{revtex4-1}

\usepackage{graphicx}
\usepackage{amsmath}
\usepackage{algorithm}
\usepackage{algorithmic}
\usepackage{media9}
\usepackage{epigraph}
\usepackage{smartdiagram}
\usepackage{verbatim}
\usepackage{lmodern}
\usepackage{listings}
\usepackage{tcolorbox}
\usepackage{color,soul}
\usepackage{listings}
\usepackage{float}

\newcommand{\commentoutA}[1]{}

\begin{document}

\preprint{LA-UR-22-32368}

\title{Graph-based Quantum Response Theory and Shadow Born-Oppenheimer Molecular Dynamics}

\author{Christian F. A. Negre}
\email{cnegre@lanl.gov}
\affiliation{Theoretical Division, Los Alamos National Laboratory, Los Alamos, New Mexico 87545}
\author{Michael E. Wall}
\affiliation{Computer, Computational, and Statistical Sciences Division, Los Alamos National Laboratory, Los Alamos, New Mexico 87545}
\author{Anders M. N. Niklasson}
\email{amn@lanl.gov}
\affiliation{Theoretical Division T-1, Los Alamos National Laboratory, Los Alamos, New Mexico 87545}

\date{\today}

\begin{abstract}
Graph-based linear scaling electronic structure theory for quantum-mechanical molecular dynamics simulations [J.\ Chem.\ Phys.\ {\bf 144}, 234101 (2016)] is adapted to the most recent shadow potential formulations of extended Lagrangian Born-Oppenheimer molecular dynamics, including fractional molecular-orbital occupation numbers [J.\ Chem.\ Phys.\ {\bf 152}, 104103 (2020); Eur.\ Phys.\ J.\ B {\bf 94}, 164 (2021)], which enables stable simulations of sensitive complex chemical systems with unsteady charge solutions. The proposed formulation includes a preconditioned Krylov subspace approximation for the integration of the extended electronic degrees of freedom, which requires quantum response calculations for electronic states with fractional occupation numbers. For the response calculations we introduce a graph-based canonical quantum perturbation theory that can be performed with the same natural parallelism and linear scaling complexity as the graph-based electronic structure calculations for the unperturbed ground state. The proposed techniques are particularly well-suited for semi-empirical electronic structure theory and the methods are demonstrated using self-consistent charge density-functional tight-binding (SCC-DFTB) theory, both for the acceleration of self-consistent field calculations and for quantum-mechanical molecular dynamics simulations. The graph-based techniques combined with the semi-empirical theory enable stable simulations of large, complex chemical systems, including tens-of-thousands of atoms. 
\end{abstract}

\maketitle

\section{Introduction}

The immense promise of linear scaling electronic structure theory \cite{SGoedecker99,DBowler12,AClark21} to study large-scale atomistic systems directly from the first principles of quantum mechanics is exceptionally difficult to realize in practice. There are a number of obstacles and problems. These challenges include:
{\em i)} a high computational pre-factor where the linear scaling benefit appears only for very large systems that in practice often are beyond acceptable time limits or available computer resources -- a challenge accentuated by highly competitive cubically scaling electronic structure solvers that have an impressive performance, especially on hybrid architectures \cite{MCawkwell12b,JFinkelstein21,JFinkelstein21B,HShang21b,cuSOLVER,BML,cublas,Mniszewski2020-pl};
{\em ii)} a reduced accuracy that often is difficult, if not impossible, to control, which may lead to instabilities and poor convergence of the self-consistent ground state solutions;
and {\em iii)} computationally expensive overheads that can limit parallel efficiency even for highly distributed calculations.
These challenges are, in particular, limiting factors for quantum-mechanical Born-Oppenheimer molecular dynamics (QMD) simulations \cite{SGoedecker94,GGalli96,ETsuchida08,FShimojo08,MCawkwell12,JVandevondele12,MArita14,FShimojo14,DOseiKuffuor14,TOtsuka16}, where all these problems coalesce, limiting the system sizes or simulation times that can be studied.
Graph-based linear scaling electronic structure theory \cite{ANiklasson16} was introduced to overcome some of these obstacles. Graph-based linear scaling QMD is a framework designed for non-metallic systems and it is particularly efficient for quasi low-dimensional structures, as for example, soft matter systems such as solvated biomolecules, where the electronic overlap between the atoms can be represented by a sparse (low-dimensional) graph \cite{ANiklasson16}. 

Graph-based linear scaling electronic structure theory \cite{ANiklasson16} with its subsequent theoretical analysis, applications, and implementations \cite{Djidjev16,MLAss18,Djidjev19,MLass20} is a very promising approach to study large atomistic systems, and combines the natural parallelism of divide-and-conquer-like
methods \cite{WYang91,PDWalker93,WYang95,IAbrikosov96,KKitaura99,SLi05,TOzaki06,YNishimoto2014,VQuan19,YNishimoto21} with the well-controlled and tunable accuracy of a numerically thresholded sparse matrix algebra \cite{S76,SPissanetzky84,bs89,YSaad96,ESchwegler96, mchallacombe97,ADaniels99,ANiklasson02,ANiklasson03,ERubensson05,DKJordan05,BAradi07,ERubensson08a,ANiklasson2011,ABuluc12,NBock13,UBorstnik14,VWeber15,PPinski15,Truflandier16,AKruchinina16}. 
It was recently considered how graph-based linear scaling electronic structure theory could be used to treat even hundreds of millions of atoms \cite{RSchade22} \footnote{Graph-based linear scaling electronic structure theory was recently used by Schade and co-workers, who call it the submatrix method or the non-orthogonal local submatrix method (NOLSM) \cite{RSchade22}, but without referring to the original and equivalent method in Ref.\ \cite{ANiklasson16}.}. Nevertheless, some shortcomings still remain. In particular, special care is required for QMD simulations of charge sensitive or reactive chemical systems, e.g.\ systems where the electronic energy gap between the Lowest Unoccupied Molecular Orbital (LUMO) and the Highest Occupied Molecular Orbital (HOMO) is opening and closing along the molecular trajectories. Electronic degeneracies, i.e.\ molecular states with equal energy at the chemical potential, and high susceptibilities associated with a low HOMO-LUMO gap can cause costly convergence and charge stability problems and difficulties calculating accurate conservative forces, which may lead to nonphysical molecular trajectories.

In this article we will combine graph-based linear scaling electronic structure theory with the most recent shadow potential formulations of extended Lagrangian Born-Oppenheimer molecular dynamics (XL-BOMD), including fractional molecular-orbital occupation numbers \cite{ANiklasson17,ANiklasson20,ANiklasson21b}. 
This shadow Born-Oppenheimer molecular dynamics framework allows QMD simulations of reactive or low-gap chemical systems that may have unsteady charge solutions  \cite{ANiklasson20}, which can be very challenging for regular direct Born-Oppenheimer molecular dynamics simulations. To achieve the necessary accuracy and stability in this shadow Born-Oppenheimer molecular dynamics formalism requires canonical quantum response calculations for electronic states with fractional occupation numbers at finite electronic temperatures. To perform these canonical response calculations we will introduce a graph-based quantum perturbation theory for systems with fractional occupation numbers. With this improved graph-based framework for QMD simulations we can then study challenging complex chemical systems using distributed computational platforms.

The prospect to use graph-based QMD to simulate chemically reactive soft matter systems is of particular interest when chemically relevant transitions appear as rare events beyond the times scales that are normally accessible to QMD simulations. Established accelerated molecular dynamics simulation methods \cite{AVoter97,AVoter02,AMD,DPerez16} 
are often difficult, if not impossible, to apply to soft matter systems that often have multiple shallow local minima where rare events are hard to detect or even to define. Nevertheless, by extending the system size it is possible to systematically enhance the probability of a chemical relevant event and thus, indirectly, accelerate the rare event dynamics. The natural parallelism, improved stability, and linear-scaling computational cost of our graph-based QMD framework makes this possible.

In this paper we will introduce graph-based quantum response calculations for systems with fractional occupation numbers. Our main motivation is to be able to perform graph-based QMD simulations using the most recent shadow potential formulations of XL-BOMD \cite{ANiklasson20,ANiklasson21b}.  
However, canonical quantum perturbation calculations performed on a partitioned graph is a very useful technique by itself. Our graph-based quantum perturbation theory presented in this article, for example, could also be used for distributed, large-scale calculations of time-independent response properties such as the magnetic susceptibility, phonon modes, electric polarizabilities, the Raman spectra, or the Born-effective charge \cite{SBaroni01,VWeber04,COchsenfeld04,JKussmann07,JKussman15}. 

The theory and the techniques will be presented for Kohn-Sham density functional theory, but our formulation is general and can also be applied to other related effective single-particle methods such as Hartree-Fock-based theory. However, the relevant system sizes for graph-based electronic structure calculations are often quite large and each separate subgraph partition may included hundreds of atoms. The graph-based method for QMD simulations is therefore particularly well-suited for semi-empirical electronic structure methods that typically are one to two-orders of magnitude faster compared to higher-level first principles methods. The new methodology will be demonstrated using self-consistent-charge density functional tight-binding theory (SCC-DFTB) \cite{MElstner98,MFinnis98,BHourahine20}, but it is also applicable to other semi-empirical methods \cite{MDewar77,MDewar85,JStewart13,CBannwarth18,PDral19,WMalone20,ZGuoqing20,CBannwarth20}. 

Our main objectives in this article are the underlying theoretical concepts and techniques for the graph-based electronic structure and response calculations that enable large-scale QMD simulations of challenging complex (non-metallic) chemical systems. 
The specific performance of our implementation on different platforms or the particular chemical behavior and mechanisms of the simulated examples, are not our priorities.

If not stated explicitly, we will assume that all operators are represented in some local atomic-orbital-like basis set. In general we will assume that these matrix representations have been orthogonalized by a congruence transformation \cite{GGolub96,ANiklasson2011} and we will use atomic units throughout the text. For the extended electronic degrees of freedom in XL-BOMD we use the electron density represented either as a continuous function or in a discretized form as a charge vector, which is particularly well-suited for SCC-DFTB theory. Generalizations to other representations of the extended electronic degrees of freedom, for example, using atomic multipoles, the density matrix, electronic wavefunctions, or the effective single-particle Hamiltonian, will not be discussed here. The underlying theoretical approaches remain the same, but some significant details in the techniques may need to be adapted to the different representations of the extended electronic degrees of freedom, e.g.\ see Ref.\ \cite{ANiklasson20b}.

The article is outlined as follows. First, we discuss the basic ideas behind general graph-based matrix function expansions and how it applies to graph-based linear scaling electronic structure theory.
Then, we present XL-BOMD in its most recent form that uses an approximate shadow Born-Oppenheimer potential energy surface. In this formulation, a kernel that determines the metric tensor of the generalized harmonic oscillator extension, appears in the equations of motion for the electronic degrees of freedom, which makes the electronic degrees of freedom oscillate around a closer estimate of the exact ground state density.
The action of this kernel in the electronic equations of motion can be updated on-the-fly using a preconditioned low-rank Krylov subspace approximation that is generated from quantum response calculations. To perform these response calculations we then introduce a time-independent graph-based quantum perturbation theory for systems with fractional occupation numbers. The graph-based canonical quantum response calculations can be partitioned into response calculations over separate subsystems, whose result can be collected for the response of the composite system. This graph-based canonical quantum perturbation theory allows for large-scale response calculations at finite electronic temperatures with a natural parallelism. We then combine the different electronic structure methods into a framework for graph-based shadow Born-Oppenheimer molecular dynamics simulations. We also present how the methodology can be used as an alternative low-complexity approach to accelerate self-consistent field iterations. The theory is then demonstrated for a few testbed examples using SCC-DFTB theory. Finally, we give a brief summary and our conclusions.

\section{Graph-based linear scaling electronic structure theory}

Graph-based linear scaling electronic structure theory is a powerful and flexible approach to achieve linear scaling complexity. The procedure combines a well-controlled and tunable accuracy 
of a thresholded sparse matrix algebra with the natural parallelism and flexibility of a divide-and-conquer-like scheme \cite{ANiklasson16}.

\begin{figure}[ht]
\includegraphics[scale=0.25]{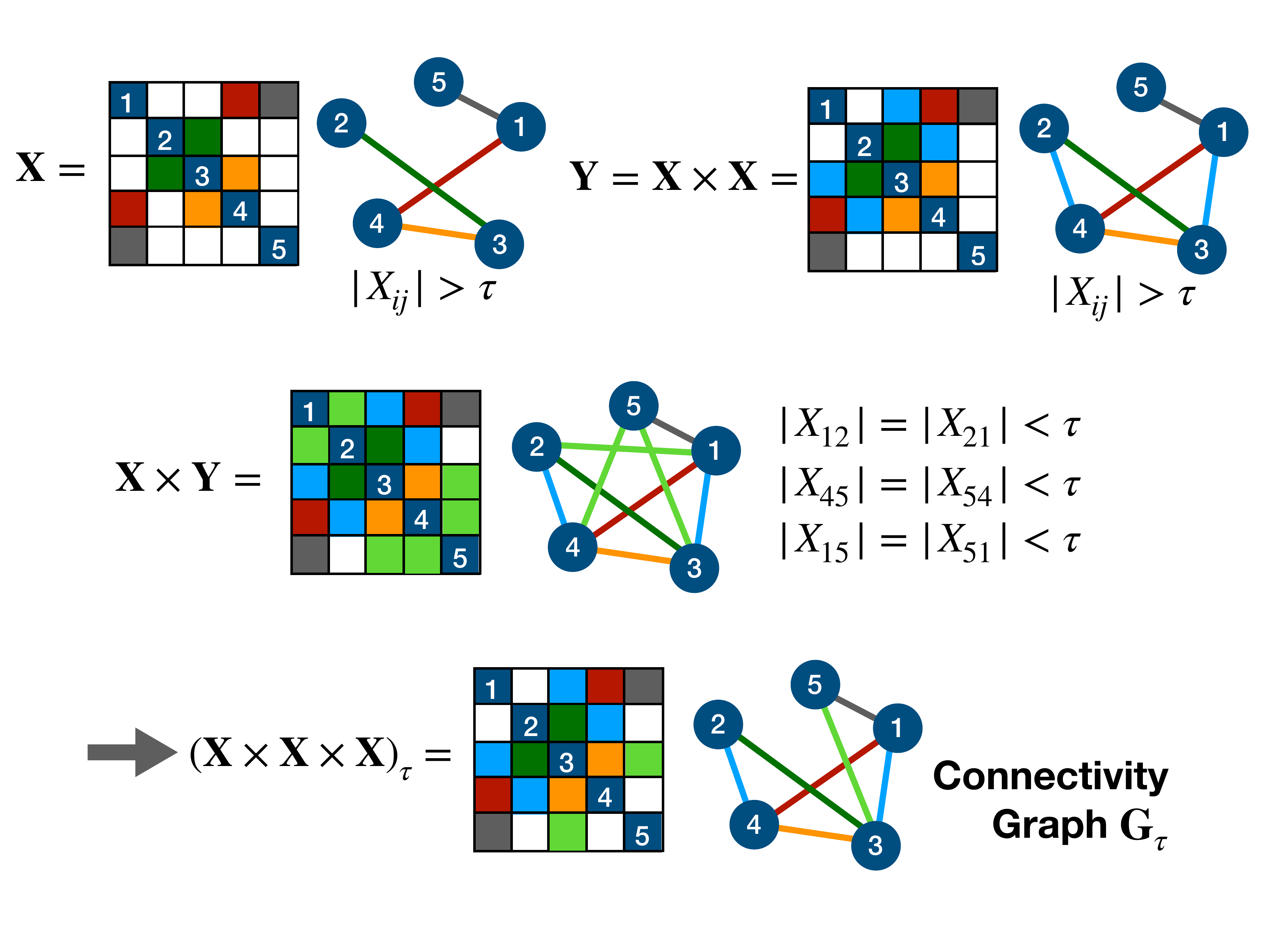}
\caption{\label{Fig_0}
{\small The step-by-step construction of an undirected symmetric connectivity (or data-dependency) graph, ${\bf G}_\tau$, using a symmetric matrix, ${\bf X}$.  The graph represent the data dependencies in a {\em globally} numerically thresholded sparse matrix function expansion, where all matrix elements are included that at any point have an absolute value above some global numerical threshold, $\tau$. White empty matrix entries corresponds to zero values.
After the second (${\bf X \times Y}$) matrix-matrix multiplication, $\vert X_{15}\vert = \vert X_{51}\vert < \tau$, but
this edge is kept in the graph, because it was previously above the threshold. The data-dependency graph, 
${\bf G}_\tau$, depicts the allowed flow of information during the polynomial expansion.}}
\end{figure}

\subsection{Numerically thresholded sparse matrix algebra}

In a matrix that is numerically thresholded, all matrix elements with an absolute value below some tunable numerical tolerance, $\tau$, are set to zero. Matrix operations with numerically thresholded matrices can often be performed with great efficiency, because only the non-zero elements above the numerical threshold have to be included in the calculations. This can often lead to a significant reduction in the computational cost, where sometimes the total number of arithmetic operations only scales linearly (or stays constant) with the matrix size \cite{SGoedecker99,Niklasson2011,DBowler12}. The numerical thresholding enables a tunable adaptive accuracy that is easy to control -- the numerical cut-off tolerance, $\tau$, is simply increased or decreased \cite{EHRubensson08}. In spirit, the numerical thresholding is similar to changing the numerical precision of the floating point representation. However, sparse matrix operations can often be problematic when they are performed in parallel on a distributed computer architecture. The necessary data transfer between nodes and a non-local data storage can lead to a large overhead.
In contrast, dense matrix operations such as matrix-matrix multiplications often can be performed with high parallel efficiency.

\subsection{Matrix polynomials on a graph}

A sparse matrix has an associated graph, where all the non-zero matrix elements are represented by edges between vertices of a graph. If we calculate a polynomial function of a matrix, ${\bf X}$, in a step-by-step process (generating each higher-order polynomial term by either multiplying by an ${\bf X}$ from the left or from the right hand-side)
we can represent the step-by-step evolution of the matrix elements with absolute values above the threshold, $\tau$, as edges 
on a graph. This process is schematically illustrated in Fig.\ \ref{Fig_0}, where the new ${\bf XY}$ term is generated by multiplying ${\bf Y}$ from the left by ${\bf X}$ before thresholding. Numerical thresholding are typically performed {\em locally} after each matrix-matrix operation. However, here we perform a {\em global} numerical threshold, where we keep all elements, whose absolute values, at any point in the polynomial expansion appear above the numerical threshold, $\tau$. This is illustrated with the $X_{15} = X_{51}$ matrix elements in Fig.\ \ref{Fig_0}, which are kept after the ${\bf X \times Y}$ multiplications even if they were below $\tau$ after the last matrix multiplication, because they were above $\tau$ in a previous step. The graph associated with the {\em globally} thresholded matrix polynomial then represents the data dependency or the data connectivity graph, ${\bf G}_\tau$, which determines how data is allowed to flow across the vertices during the polynomial matrix expansion and only the non-zero matrix entries corresponding to ${\bf G}_\tau$ are used during a polynomial expansion. 

A globally thresholded matrix polynomial
is thus given by the step-by-step polynomial matrix expansion, where we only include matrix elements corresponding to the vertices and edges
of the data-dependency graph, ${\bf G}_\tau$, after each matrix-matrix operation. This procedure, where we use a fixed sparsity pattern determined by the graph ${\bf G}_\tau$, we refer to as {\em matrix polynomial on a graph} -- denoted, for example, by $f_{G}({\bf X})$ or $f({\bf X}) \vert_G $. 

For the matrix operations on a graph it is important to remove any fill-in elements that don't belong to the graph after each individual matrix-matrix multiplication in step-by-step procedure after each matrix multiplication \cite{ANiklasson16}. For example, we do not perform the threshold after ${\bf Y}_2 = {\bf X}^3$, but instead already for the intermediate operation ${\bf Y}_1 = {\bf X} \times {\bf X}$ and then again for ${\bf Y}_2 = {\bf X}\times {\bf Y}_1$ (here with a left-hand side multiplication), as illustrated in Fig.\ \ref{Fig_0}. Only with this incremental step-by-step procedure can we cover all possible data dependencies that may appear in the polynomial matrix expansion. The example in  Fig.\ \ref{Fig_0} is for a {\em symmetric} matrix polynomial and a symmetric data-connectivity graph. Generalizations to matrix operations on non-symmetric directed data-dependency graphs are straightforward (see supporting information).

Instead of identifying the data dependency graph from a global numerical threshold of a matrix polynomial, we can, alternatively, simply chose any data-connectivity graph, ${\bf G}$, that then limits the allowed data flow during a polynomial matrix expansion on this graph.
A matrix polynomial calculated on this graph is given by the same incremental step-by-step expansion, where only matrix elements corresponding to the graph ${\bf G}$ are kept after each increase in the polynomial order. Examples are given in the supporting information. 
The advantage of using a graph determined by a global numerical threshold is that we can then easily control and tune the accuracy of the matrix function by increasing or decreasing the threshold, $\tau$. This is not possible when we chose the graph in some arbitrary way.

A graph ${\bf G}$ can be partitioned into subgraphs consisting of non-overlapping core (c) parts and overlapping halos (h) (see Fig.\ \ref{Fig_1}). For example, we can chose one subgraph for each separate vertex of ${\bf G}$, including all its edges and its directly connected neighboring vertices. This partitioning forms a set of overlapping subgraphs. The non-overlapping vertex of a graph is its core part and the overlapping shared edges and vertices belong to the halo.

There is an important and useful one-to-one relation between a sparse matrix function expansion on a graph ${\bf G}$, which may have been determined from a global numerical threshold or just chosen arbitrarily, and a collection of separate matrix function expansions using dense matrix algebra over a set of smaller principal submatrices, where the submatrices are determined by the overlapping subgraphs from the partitioning of the data dependency graph, ${\bf G}$.  
This observation forms the basis of graph-based linear scaling electronic structure theory \cite{ANiklasson16}. 
The equivalence enables a natural parallelism in electronic structure calculations, similar to a divide-and-conquer approach, but with the tunable  accuracy of a numerically thresholded sparse matrix algebra. 
After the initial formulation of graph-based linear scaling electronic structure theory a more rigorous mathematical analysis was subsequently given in Refs.\ \cite{Djidjev16,Djidjev19}.

\subsection{One-to-one relation between a sparse matrix function and collected small dense matrix functions}

\begin{figure}[ht]
\includegraphics[scale=0.27]{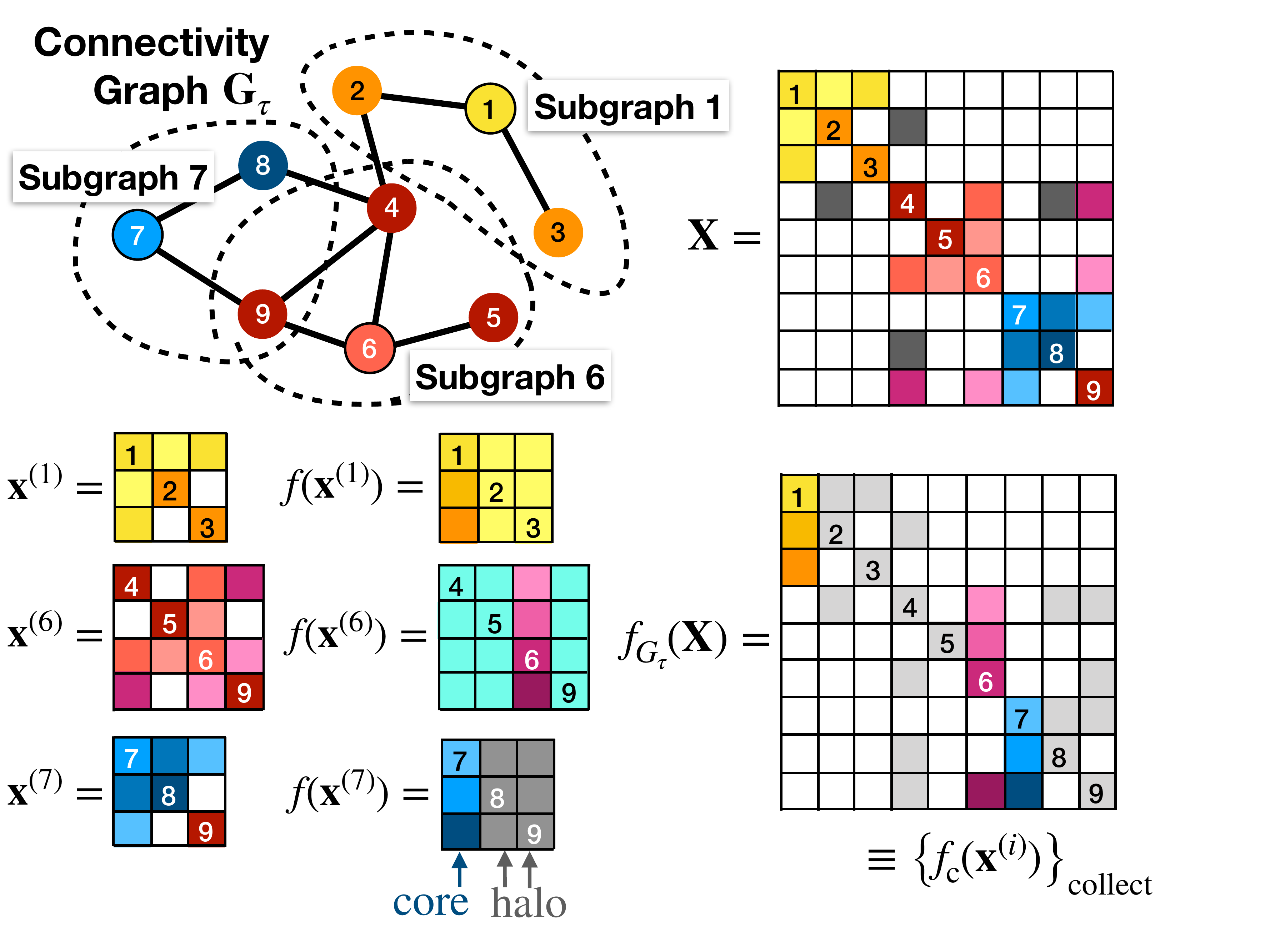}
\caption{\label{Fig_1}
{\small The connectivity graph, ${\bf G}_\tau$, (here undirected and symmetric) represents the data dependencies in the {\em globally} numerically thresholded sparse matrix function expansion, ${f}_{G_\tau}({\bf X}) = \sum_k c_k T_k({\bf X})\big\vert_{G_\tau}$, with a threshold $\tau$
(See Fig.\ \ref{Fig_0}). 
The partitioned subgraphs of the different core vertices (e.g.\ the partial subset 1, 6, and 7) defines the principal submatrices
(e.g.\ ${\bf x}^{(1)}$, ${\bf x}^{(6)}$, and ${\bf x}^{(7)}$) that are extracted from ${\bf X}$ (here symmetric).  White empty matrix entries corresponds to zero matrix values. Separate {\em dense} matrix function expansions are then performed over these submatrices (e.g.\ the partial subset ${f}({\bf x}^{(1)})$, ${f}({\bf x}^{(6)})$, and ${f}({\bf x}^{(7)})$. The results of the separate matrix function expansions are then to be collected column by column (or row by row) from their corresponding core-column parts $\{f_{\rm c}({\bf x}^{(i)})\}$, which gives the equivalent result to the globally thresholded
matrix function expansion, i.e.\ ${f}_{G_\tau}({\bf X}) = \left\{ {f}_{\rm c}({\bf x^{(i)}})\right\}_{\rm collect}$ where $f({\bf X})$ is expanded on ${\bf G}_\tau$ \cite{ANiklasson16}. The core-column parts are, for example, column 1 for ${f}({\bf x}^{(1)})$, column 3 for ${f}({\bf x}^{(6)})$ and column 1 for ${f}({\bf x}^{(7)})$. The partially collected matrix is illustrated in the lower right, indicating in grey the remaining elements to be obtained from other subgraphs.  Generalizations to sets of non-symmetric and non-commuting matrices using directed data-dependency graphs, ${\bf G}_\tau$, are straightforward (see pseudo-code in the supporting information)}.}
\end{figure}

In graph-based linear scaling electronic structure theory, a matrix function calculated on a graph,
\begin{equation}\label{f_tau}
f_G({\bf X}) =  \sum_k a_k T_k({\bf X}) \big\vert_{G}~,
\end{equation}
i.e.\ where the flow of data is limited by some  graph, ${\bf G}$,
is equivalent to the concatenated results of the core-centered columns of the same matrix polynomial performed over a set of separate dense principal submatrices, $\{{\bf x}^{(i)}\}$, such that
\begin{equation}\label{sub_f}
f_G({\bf X}) =  \left\{ f_{\rm c}({\bf x}^{(i)})\right\}_{\rm collect}. 
\end{equation}
Here the curly brackets, $\{ \cdot \}_{\rm collect}$,
denote the operation of {\em collecting} the columns (or rows) of the submatrix function corresponding to matrix elements connecting to the core parts of the subgraphs, $f_{\rm c}({\bf x}^{(i)})$, into the composite matrix function, $f_G({\bf X})$, calculated on ${\bf G}$ (See Fig.\ \ref{Fig_1}). Each core column includes matrix elements of the core and its overlap with the halo. $T_k({\bf X})$ in Eq.\ (\ref{f_tau}) are assumed to be basis polynomials of various orders with coefficients $a_k$. 

Notice that the data-connectivity graph ${\bf G}$ does not have to be determined by the global numerical threshold of $f({\bf X})$. Any sparse graph can be chosen. The relation in Eq.\ (\ref{sub_f}) still holds (see supporting information). This freedom provides a lot of flexibility in how we can control the accuracy and improve the computational efficiency. In particular, we can always add edges to the graph without losing accuracy.

The relation between the matrix function on the graph and the collected submatrix polynomials is illustrated in Fig.\ \ref{Fig_1}. The partitioning of the data dependency graph, ${\bf G}_\tau$, into subgraphs determines the principal submatrices, ${\bf x}^{(i)}$. Each subgraph and corresponding submatrix has a core and a halo (h). For example, in Subgraph 1 in Fig.\ \ref{Fig_1} the vertex 1 (yellow) is in the core and vertices 2 and 3 (orange) are in the halo. Here each core is chosen to be a single vertex (corresponding to a diagonal matrix element) and only the subgraphs for vertex 1, 6, and 7 are shown out of a total of 9 subgraphs. 
The principal submatrices, ${\bf x}^{(i)}$, may have matrix elements that are not present in the connectivity graph corresponding to matrix elements, $X_{ij}$, that are initially 0 (empty white matrix entries), but which become finite at some point during the matrix function expansion. An example is the edge between node 2 and 3 in subgraph 1 in Fig.\ \ref{Fig_1}. The separate results of the matrix elements of the core columns of the dense submatrix function expansions are then extracted and put together column by column (or row by row) to assemble the matrix function expansion of the full sparse matrix that otherwise would be calculated on the data connectivity graph, ${\bf G}_\tau$, or with the corresponding global numerical threshold \cite{ANiklasson16,MLAss18,MLass20}. This is the ``collect'' operation, as indicated by the subscript on the very right-hand side of Eq.\ \ref{sub_f} and in the lower right corner of Fig.\ \ref{Fig_1}. 

The one-to-one relation between a matrix function on a graph and its collected submatrix functions holds also for a polynomial matrix expansion using a variety of non-commuting and non-symmetric matrices, including directed (non-symmetric) data-dependency graphs, ${\bf G}$. In these more general cases some special care is required to account for the allowed data flow \cite{ANiklasson16} (see supporting information). 
For a matrix expansion of non-commuting or non-symmetric matrices on a general directed data dependency graph, the principal submatrices are extracted either by columns or by rows. The higher-order polynomial terms on the graph are then built by multiplying matrices either from the left-hand side (for the column case) or from the right-hand side (for the row case) in the incremental step-by-step polynomial expansion. Only matrix elements on the graph are kept after each matrix operation. This matrix polynomial on the graph is then equivalent to the collected set of the submatrix polynomials that are collected either by columns or by rows from their core parts. Examples that illustrate the separate column and row-wise versions for the non-commuting or non-symmetric polynomial expansions on a graph are provided in the supporting information. 

\subsection{Fermi-operator expansion on a graph}

In graph-based linear scaling electronic structure calculations, the matrix function, $f_G({\bf X})$, can be a Fermi-operator
expansion, which generates the effective single-particle density matrix that determines the charge density (see Eq.\ (\ref{rho_D}) below), or the inverse factorization of the overlap matrix \cite{ANiklasson16,CNegre16,MLAss18,MLass20}, which is used to transform the generalized eigenvalue problem into a regular orthonormalized form. 
These operations can all be expressed as polynomial functions of the Kohn-Sham Hamiltonian or the overlap matrix \cite{ANiklasson2011,CNegre16}.
For example, when we calculate the effective single-particle density matrix, ${\bf D}$, from the Kohn-Sham Hamiltonian matrix, ${\bf H}$,
we can do so with a Fermi operator expansion on a graph.
In this case
\begin{align}
{\bf D}_{G_\tau} &=  \left[e^{\beta ({\bf H}-\mu I)}+{\bf I} \right]^{-1} \bigg\vert_{G_\tau} \label{Fermi}\\
& = \sum_k a_k T_k({\bf H}) \big\vert_{G_\tau},
\end{align}
where ${\bf I}$ is the identity matrix,
$\{a_k\}$ some set of coefficients and $\{T_k({\bf H})\}$ is a set of polynomials, e.g.\ Chebychev polynomials of various orders,
$\beta = 1/(k_BT)$ is the inverse electronic temperature, and $\mu$ is the chemical potential. 
We have here assumed that ${\bf H}$ is in an orthogonalized matrix representation \cite{ANiklasson2011,DBowler12}, but generalizations to non-orthogonal expansions are straightforward \cite{ANiklasson05,ANiklasson16,2016progress,BML,LATTE}. 

In a polynomial Fermi-operator expansion on the data-dependency graph, ${\bf G}_\tau$, all matrix elements outside of ${\bf G}_\tau$ are removed in each incremental step-by-step calculation of the polynomials \cite{ANiklasson16}, in the same way as for a global numerical thresholding illustrated in Fig.\ \ref{Fig_0}.
However, the dense matrix function expansions of the principal submatrices, $f_c({\bf x^{(i)}})$ in Eq.\ (\ref{sub_f}), defined by the subgraph partitioning of ${\bf G}_\tau$, are agnostic to how the functions are calculated,
because no graph or numerical threshold is invovled. The Fermi operator expansion of the principal submatrices, ${\bf h}^{(i)}$, extracted from ${\bf H}$ that are
determined by the graph partitioning of ${\bf G}_\tau$, can therefore be performed in any manner \cite{ANiklasson16}. In particular, in the limit of vanishing electronic temperature ($\beta \rightarrow \infty$) we can use
a fast recursive Fermi-operator expansion method, where
\begin{align}
{\bf d}^{(i)} &= \Theta \left(\mu {\bf I}^{(i)} - {\bf h}^{(i)}\right) \label{Step} \\
 &= \lim_{n \rightarrow \infty} g_n(g_{n-1}(\ldots g_0({\bf h}^{(i)})\ldots )), \label{Recur}\\
{\bf D}_{G_\tau} & = \left\{{\bf d}^{(i)}_{\rm c}\right\}_{\rm collect}. \label{dCollect}
\end{align}
Here $\Theta(\cdot)$ is the Heaviside step function and $\{g_n({\bf x}^{(i)})\}$ is some set of purification or spectral projection polynomials \cite{RMcWeeny56,APalser98,ADaniels99,KNemeth00,ANiklasson02,ANiklasson03,ANiklasson03B,DKJordan05,ANiklasson2011,ERudberg11,EHRubensson11,DBowler12,Truflandier16,PSuryanarayana13,EHRubensson14} 
that rapidly reaches convergence for the idempotent subgraph density matrices, $\{{\bf d}^{(i)}\}$. Alternatively, we may formulate the recursive sequence in Eq.\ (\ref{Recur}) as a convolutional deep neural network \cite{JFinkelstein21,JFinkelstein21B}.
The globally thresholded density matrix, ${\bf D}_{G_\tau}$, is then assembled by the core parts (c) of the subgraph density matrices, $\{{\bf d}^{(i)}_{\rm c}\}_{\rm collect}$.
This relation, as in Eq.\ (\ref{dCollect}), is of critical importance to any operations we perform in graph-based electronic structure theory as well as to the graph-based quantum perturbation theory presented later in this article.
The separate recursive expansions (or purifications) over the smaller separate dense submatrices can be performed with close
to peak performance on specialized hardware such as many-core processors  \cite{HShang21a,HShang21b}, graphics processing units (GPUs) \cite{MCawkwell12b,ANiklasson16,RSchade22}, and artificial-intelligence (AI) accelerating mixed-precision Tensor cores or Tensor processing units (TPUs) \cite{JFinkelstein21,JFinkelstein21B,RSchade22,JFinkelstein22,RPederson22}.

Several alternatives to the recursive expansion approach in Eq.\ (\ref{Recur}) can be used, including expansions in the diagonal molecular-orbital (or eigenbasis) representation, where the molecular orbitals are given from solutions of the corresponding quantum mechanical eigenvalue problem.
In the general non-orthogonal case \cite{ANiklasson05,ANiklasson16}, the Fermi-operator expansion on the graph then corresponds to the solution of a set of small generalized Kohn-Sham eigenvalue problems,
one for each subgraph, $i$, and corresponding principal submatrix ${\bf h}^{(i)}$ of the full composite Kohn-Sham Hamiltonian, ${\bf H}$. The composite sparse density matrix, ${\bf D}_{G_\tau}$, is then collected from the core parts of each subgraph density matrix, ${\bf d}^{(i)}$, given in a molecular-orbital representation. 
As in  Eq.\ (\ref{dCollect}), the density matrix, ${\bf D}_{G_\tau}$, of the composite system is then given by the collected core-column parts of the subgraph density matrices, i.e.\
\begin{align}
&{\bf D}_{G_\tau}(\mu) = \left\{ {\bf d}^{(i)}_{\rm c}(\mu)\right\}_{\rm collect} \label{DD0} \\
& ~~~~~~ = \left\{ \sum_k \left. \left(e^{\beta(\epsilon_k^{(i)} - \mu)}+1\right)^{-1}{\bf v}^{(i)}_k {{\bf v}^{(i)}_k}^T\right \vert_{\rm c} \right\}_{\rm collect}. \label{DD}
\end{align}
Here $\{{\bf v}^{(i)}_k\}$ are the molecular-orbital eigenvectors of the subgraph Hamiltonian principal submatrices, ${\bf h}^{(i)}$, in their orthogonalized form, where
\begin{align}
&{\bf h}^{(i)} {\bf v}^{(i)}_k = \epsilon_k^{(i)} {\bf v}_k^{(i)} \label{SmallKS}\\
&{\bf h}^{(i)} = {{\bf z}^{(i)}}^T{\bf h}_{\rm ao}^{(i)}{\bf z}^{(i)}, ~~ {{\bf z}^{(i)}}^T{\bf s}^{(i)} {\bf z}^{(i)} = {\bf I}^{(i)}.\label{ksSmall}
\end{align}
${\bf s}^{(i)}$ are the principal submatrices of the overlap matrix for the full composite system, ${\bf S}$, and 
${\bf z}^{(i)}$ is some inverse factorization of ${\bf s}^{(i)}$. The subsystem Hamiltonian in the non-orthogonal
atomic-orbital (ao) representation is given by ${\bf h}_{\rm ao}^{(i)}$. These have been extracted as principal submatrices from the effective single particle Hamiltonian, ${\bf H}_{\rm ao}$, using some estimated data-dependency graph, ${\bf G}_\tau = {\bf G}_\tau^{\rm ao}$, for the atomic orbital representation. Notice that only the core-column parts (c) of the eigenvector outer product in Eq.\ (\ref{DD}) are included in the column-by-column (or row-by-row) assembly of ${\bf D}_{G_\tau}$. The density matrix can also be assembled in the non-orthogonal atomic-orbital representation, where
\begin{equation} \begin{array}{l}
    {\bf d}^{(i)}_{\rm ao,c} = {\bf z}^{(i)} {\bf d}^{(i)}_{\rm c} {{\bf z}^{(i)}}^T, \\
    {\bf D}_{{G_\tau}}^{\rm ao}(\mu) = \left\{ {\bf d}^{(i)}_{\rm ao,c}(\mu)\right\}_{\rm collect}.
    \end{array}
\end{equation}
Apart from the congruence transformations with ${\bf z}^{(i)}$, the non-orthogonal problem is thus equivalent to the orthogonal case.

\subsection{Estimating the connectivity graph}\label{Estimate_G}

The data connectivity graph, ${\bf G}_\tau$, given from a global numerical threshold, $\tau$, of some matrix function, is in general not known {\em a priori}.
In fact, a global numerical threshold of a matrix function requires that we first calculate the full matrix function before it can be partitioned into separate submatrix calculations. This may seem like a highly unpractical approach.
However, in graph-based QMD simulations, or in iterative self-consistent field optimizations, the data connectivity graph, ${\bf G}_\tau$, can be estimated from previous steps, including redundant estimates from new connections formed as the atoms move or the electronic structure changes \cite{ANiklasson16}.
Our favourite choice is to estimate the data dependency graph from the non-zero structure of the numerically thresholded density matrix (either in the orthogonalized or the atomic-orbital representation) from a previous self-consistent-field (SCF) iteration or molecular dynamics time step, including any new matrix elements that may have formed in the thresholded Hamiltonian. Additional connectivity is then added by performing paths of length two of this graph. After performing the Fermi-operator expansions on the partitioned subgraphs, the collected density matrix is then, once again, numerically thresholded and the process is repeated. This technique allows a rapid adaptation to new data dependency flows and sparsity patterns in combination with an easily numerically tunable accuracy \cite{ANiklasson16}.
This method will also be used in the examples of our graph-based QMD scheme.

Alternative approaches include estimates from the thresholded envelope functions of the occupied eigenvectors, i.e.\ where we take the absolute values of elements of each eigenvector of the subsystems, $\{\vert {\bf v}_k^{(i)}\vert\}$. These can then be used to construct an envelope (E) density matrix, ${\bf D}_{\rm E}$, for the full composite system. In this way decoherence or cancellations in the superposition of the fluctuating tails of the eigenstates, which may lead to a localization of the regular density matrix, are avoided. New data dependency connections may then be included from the new thresholded Hamiltonian, where the new data connectivity graph, ${\bf G}$, is estimated from
\begin{equation}
    {\bf D}_{\rm E}{\bf H}_{\rm new} + {\bf H}_{\rm new} {\bf D}_{\rm E} \rightarrow {\bf G}.
\end{equation}
Other more {\em ad hoc} alternatives include, for example, using some power, $m$, of the thresholded overlap matrix, ${\bf S}^m$, or the Hamiltonian, ${\bf H}^m$, to estimate the data connectivity graph. For many systems with a significant HOMO-LUMO gap the matrix square, i.e.\ using $m=2$, is often sufficient. However, this {\em ad hoc} approach does not account for changes in the electronic overlap present in the density matrix and can therefore lead to uncontrolled errors.

An important feature of the different estimates of the data connectivity graph is that, even if we chose to have a fixed core part, the halo is fluctuating in size after each MD time step or SCF iteration \cite{ANiklasson16}. This can lead to some limitations in designing the calculations.

\subsection{Choosing the graph partitioning}

The flexibility in determining the connectivity graph, ${\bf G}_\tau$ extends to the partitioning of the subgraphs into the core and the halo, which in turn determine the principal submatrices. In particular the core is not limited to the case of a single matrix element illustrated in Fig.~\ref{Fig_1}. The computational efficiency can be increased by including multiple matrix elements in the core while keeping the combined number of halo elements per core element small \cite{ANiklasson16}.
Highly efficient core partitionings can be constructed using off-the-shelf graph methods \cite{ANiklasson16,MLass20,metis,metissource}
that identify communities of closely connected vertices. Such graph partitioning, or other alternative partitioning methods \cite{RSchade22}, have been shown to decrease the average proportion of halo vertices in the subgraphs leading to a  significant reduction in the computational overhead  \cite{ANiklasson16,Djidjev19,MLass20,RSchade22}. 

\begin{figure}[ht]
  \includegraphics[width=0.5\textwidth]{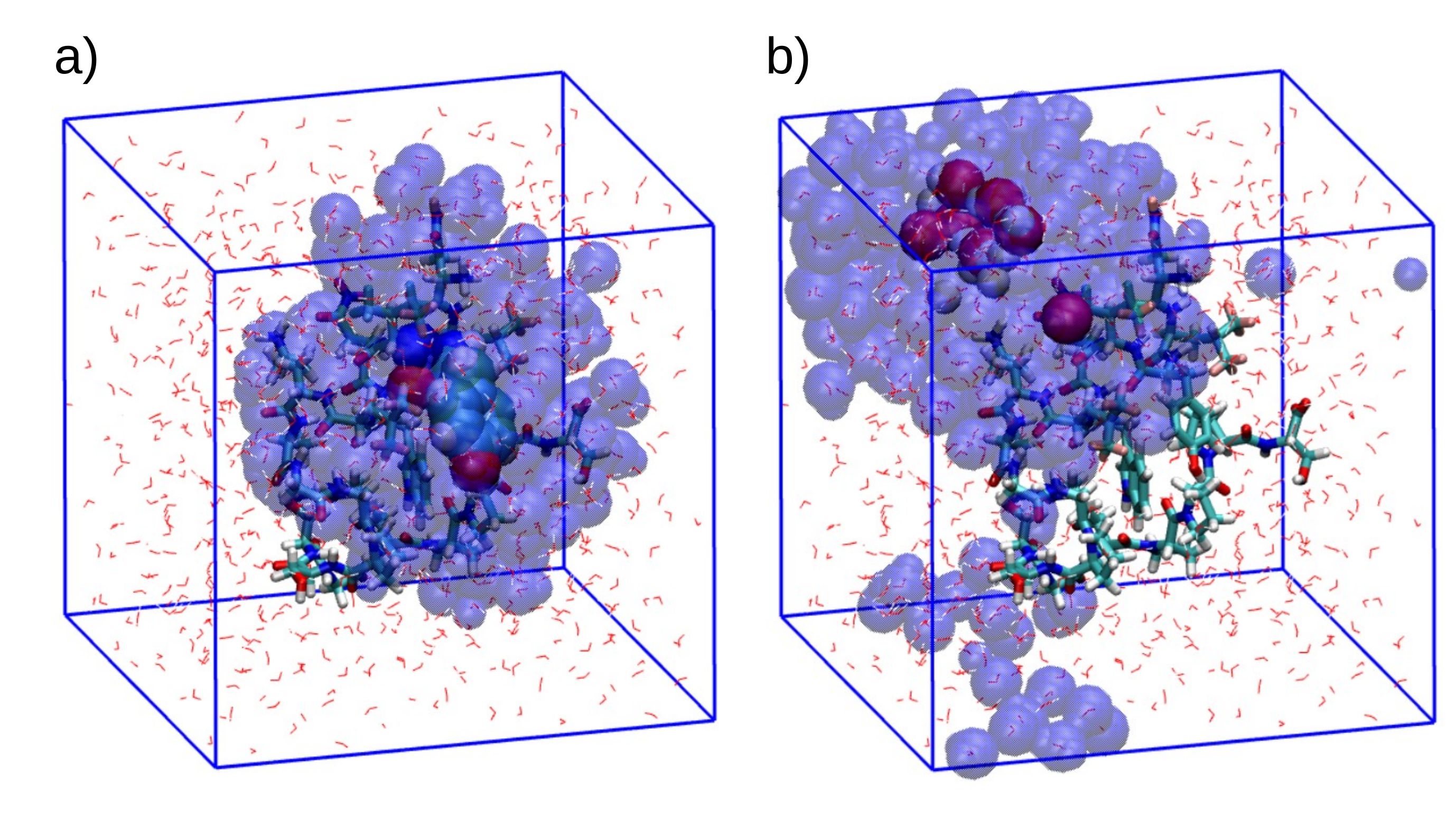}
  \caption{Chemical representation for two different subgraphs of the Trp-cage protein system solvated in water in a periodic box \cite{Neidigh2002-su}. The protein and water solvent are shown using a licorice and stick representation respectively. Carbon, oxygen, nitrogen, and hydrogen are depicted with magenta, read, blue, and white colors respectivelly.  The METIS graph-partitioning algorithm was used to partition the graph \cite{metis,metissource}. a) A small tyrosine residue is here automatically selected as the core part for one of the subgraphs (shown in VDW representation) with the halo atoms shown in translucent blue. b) A small tightly connected water cluster is automatically selected as the core (shown in VDW representation) together with the halo atoms in translucent blue. Halo atoms that appear distant from each other are actually nearby, due to the periodic boundary conditions.}
  \label{Tr_Cage}
\end{figure}

It is important to note that the data connectivity graph guided by the range of the thresholded density matrix is governed mainly by the electronic overlap of the molecular orbitals and not by the interatomic distances, e.g.\ based on some radial cutoff. This can be seen in the example of a solvated Trp-cage protein structure shown in Fig.\ \ref{Tr_Cage}, where the halo atoms surrounding a core region can be quite non-uniform.  The data connectivity graph therefore gives a picture of the electronic overlap or the chemical bonding in the same way as the numerically thresholded density matrix, ${\bf D}_{G_\tau}$. This also means that graph-based QMD will be inefficient for metallic systems, where we have itinerant electronic states that, in practice, overlap with all atoms. The graph-based calculations will still work correctly, but the lack of sparsity will lead to a single subgraph
that encloses the whole system. The natural parallelism and reduced linear scaling complexity is then lost.

\begin{figure}[ht]
  \includegraphics[width=0.5\textwidth]{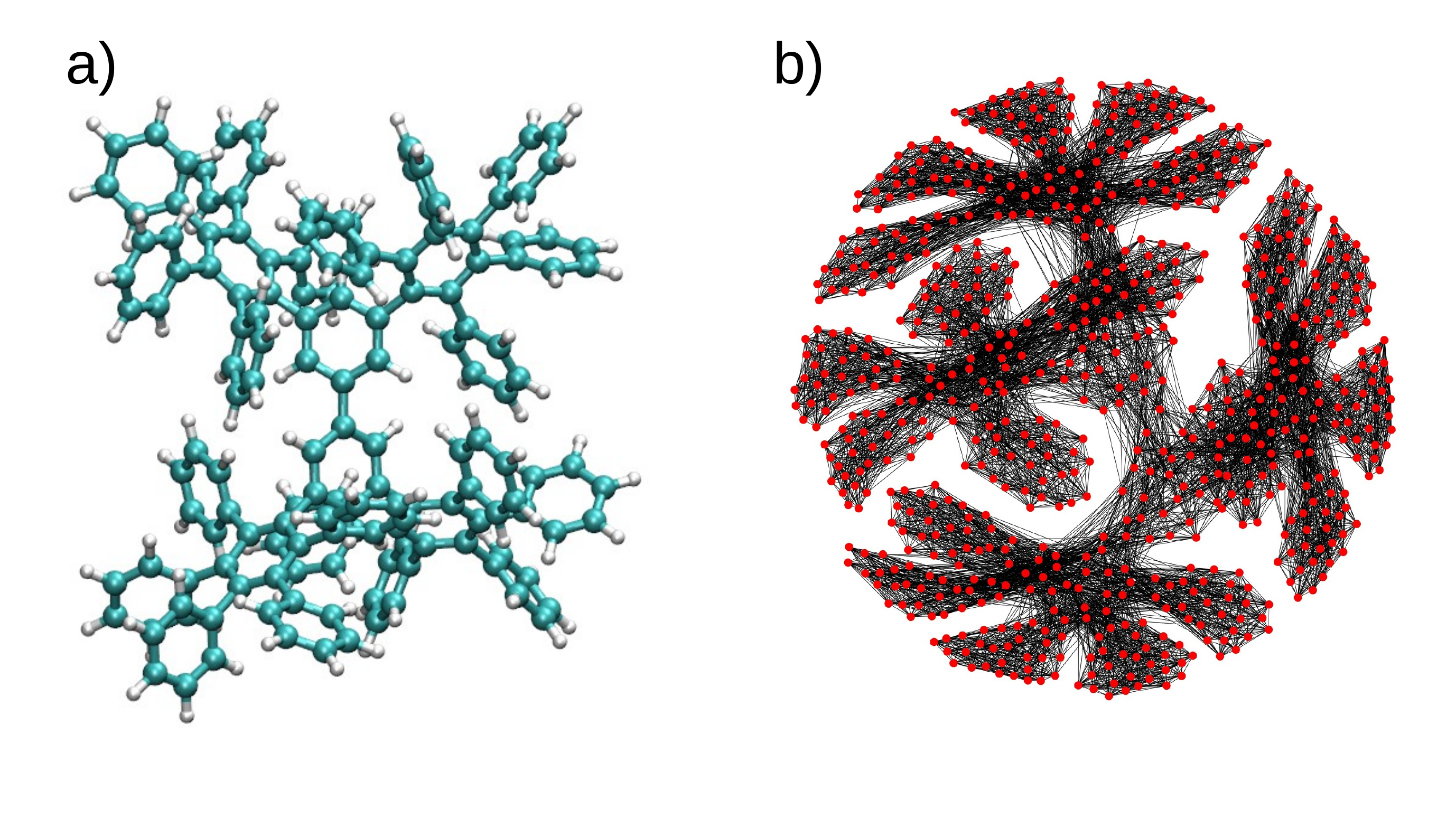}
  \caption{a) Chemical representation of a phenyl dendrimer system.   Carbon  and  hydrogen  atoms  are depicted  in  cyan  and  white  spheres  respectively \cite{Djidjev2019-mv,Ghale2017-xc}. b) Original graph partitioning extracted from the effective single-particle density matrix of the phenyl dendrimer molecular structure using the METIS graph-partitioning algorithm \cite{metis,metissource}. Note the fractal-like structure of the graph and the formation of natural communities.}  
  \label{Graphpartitioning}
\end{figure}

 Partitioning the graph by the density matrix will naturally divide soft matter systems into chemically relevant units that will often correspond to a chemically sound division. As an example of the latter, in Fig.\ \ref{Tr_Cage}, we show the partitioning result applied to a system composed of a Trp-cage miniprotein (a 20 residues peptide) solvated in water, which is an archetypal model for a soft matter system. By partitioning the connectivity graph for the merged graphs of the density matrix and the corresponding Hamiltonian matrix, using an off-the-shelf community detection algorithm \cite{metis},  we can see how key parts can be revealed. The left panel of Fig.\ \ref{Tr_Cage} shows a tyrosine residue that automatically has been captured as a core part. We have added the surrounding halo atoms in translucent blue to show the extent of the connectivity. 
The right panel of Fig.\ \ref{Tr_Cage} shows how a closely connected water cluster is captured within a core partitioning. The ability of graph-partitioning the data-connectivity graph determined by the electronic overlap on identifying chemically intuitive fragments could be of great interest as an automatic analysis tool for large complex simulations.  In particular if fragments appear in core partitions that are consistent over time.

Another example of an automatic graph clustering is shown in Fig.\ \ref{Graphpartitioning}, where a METIS community detection algorithm \cite{metis,metissource} was used to identify the core parts of a graph 
corresponding to a phenyl dendrimer system. The left panel shows the molecular structure and the right panel shows the clustered graph with the more closely interconnected communities. This automatic core partitioning can then be used in the graph-based electronic structure calculations to reduce the computational overhead.

\subsection{Newton-Raphson scheme to find the chemical potential}

To study chemical systems where the electronic HOMO-LUMO gap is small or vanishing it is necessary to avoid instabilities. This is of particular importance in QMD simulations, where discontinuous crossings of nearly degenerate states may appear almost instantaneously. The problems can be avoided if we introduce fractional occupation numbers, i.e.\ we use a temperature-dependent free energy formulation for the electronic states \cite{NMermin63,NMermin65,RParr89}. Using fractional occupation numbers leads to some additional challenges in our graph-based approach to electronic structure calculations. For example, it is no longer sufficient to determine the shared chemical potential $\mu$ in Eq.\ (\ref{Fermi}) or (\ref{Step}) over all the subgraph systems {\em somewhere} in the shared HOMO-LUMO gap \cite{ANiklasson16,EHRubensson14}.
Instead we need the exact shared value of the chemical potential, $\mu$. This can be achieved using an iterative Newton-Raphson scheme as presented in Ref.\ \cite{ANiklasson08b}, which rapidly determines the chemical potential that gives the correct total occupation for a system with fractional occupation numbers.

In the Newton-Raphson scheme we assume that we have calculated the $\mu$-dependent effective single-particle density matrices, ${\bf d}^{(i)}_{\rm c}(\mu)$, of each subsystem (core + halo) from the Fermi function with some initial guess of $\mu$ using the diagonalization as in Eqs.\ (\ref{DD0})-(\ref{ksSmall}).
The correct chemical potential, $\mu$, which is shared over all the subsystems, is determined from the occupation condition 
of the composite system, i.e.\
\begin{align}
{\rm Tr}\left[{\bf D}_{G_\tau}(\mu) \right] - N_{\rm occ} = 0 .
\end{align}
With some sufficiently good initial guess of $\mu$ this nonlinear equation can be solved iteratively using Newton-Raphson's method \cite{ANiklasson08b}, where
\begin{equation}\begin{array}{ll}
& \mu_{n+1} = \mu_n - \left( \beta{\rm Tr}\left[{\bf D}_{G_\tau}^{(n)}({\bf I} - {\bf D}_{G_\tau}^{(n)}) \right]\right)^{-1}\\
&  ~~~~~~~~~~~~~~~~~~~~~ \times \left({\rm Tr}\left[{\bf D}_{G_\tau}^{(n)} \right] - N_{\rm occ}\right).
\end{array}\label{NewtonRaphson}
\end{equation}
Here we have used the notation ${\bf D}_{G_\tau}^{(n)} \equiv {\bf D}_{G_\tau}(\mu_n)$.
The different terms in each Newton-Raphson iteration are given from the separate subsystems based on the relation in Eqs.\ (\ref{DD0}) and (\ref{DD}), where
\begin{equation} \begin{array}{ll}
&{\rm Tr}\left[{\bf D}_{G_\tau}^{(n)} \right] = {\rm Tr} \left\{{\bf d}^{(i)}_{\rm c}\right\}_{\rm collect}. \label{d_Collect} \\
&~~~~~~~  = \sum_i {\rm Tr} \left[ \sum_k \left. \left(e^{\beta(\epsilon_k^{(i)} - \mu_n)}+1\right)^{-1}{\bf v}^{(i)}_k {{\bf v}^{(i)}_k}^T \right \vert_{\rm c} \right]\\
&~~~~~~~  = \sum_{i,k,j\in {\rm c}} f^{(i)}_k(\mu_n) 
\left({v}^{(i)}_{k,j}  \right)^2 
\end{array}
\end{equation}
and
\begin{equation}\begin{array}{ll}
& \left( \beta{\rm Tr}\left[{\bf D}_{G_\tau}^{(n)}({\bf I} - {\bf D}_{G_\tau}^{(n)}) \right]\right)^{-1} \\
&~~~
=\left(\beta\sum_{i,k, j \in c} f^{(i)}_k(\mu_n)\left( 1-f^{(i)}_k(\mu_n)\right)\left({v}^{(i)}_{k,j}  \right)^2 \right)^{-1},
\end{array}
\end{equation}
with the Fermi occupation factors
\begin{align}
&f^{(i)}_k(\mu_n) = \left[e^{\beta\left(\epsilon_k^{(i)} - \mu_n\right)}+1\right]^{-1}.
\end{align}
From each separate subsystem, $i$, we need only the sum of the square of the vector elements, $j$, of the eigenvectors, $\{{\bf v}_k^{(i)}\}$, belonging to the
core part, i.e.\ $\{({v}^{(i)}_{k,j\in {\rm c}})^2\}$, for each eigenstate $k$ and the eigenvalues $\{\epsilon^{(i)}_k\}$. This
makes the Newton-Raphson scheme straightfoward to parallelize efficiently.

Netwon-Raphson's method is quadratically convergent. In QMD simulations or in self-consistent field iterations the chemical potential from previous step can be used as an initial guess. It can also be propagated or extrapolated from a sequence of previous time steps. Often only a few iterations are then necessary to reach a tightly converged value of the shared chemical potential.

The Newton-Raphson scheme in Eq.\ (\ref{NewtonRaphson}) illustrates the general power of the graph-based approach to electronic structure calculations with its equivalence between the matrix function on a graph and its collected matrix functions of the principal submatrices, which described by the graph partitioning construction in Eq.\ (\ref{dCollect}). Most algorithms for a full composite sparse system can be partitioned into operations over independent subsystems, which allows for a natural parallelism.

The Newton-Raphson scheme in Eq.\ (\ref{NewtonRaphson}) is efficient for systems with fractional occupation numbers and a small HOMO-LUMO gap. For systems with integer occupation of the electronic states and a finite HOMO-LUMO gap, there are more efficient alternative techniques that can be used even in combination with recursive Fermi-operator expansion methods \cite{ANiklasson16,EHRubensson14}.

\section{XL-BOMD}

In graph-based QMD simulations the overhead from the required self-consistent electronic ground state optimization, which is required prior to each force evaluation, can be reduced by using modern Car-Parrinello-like methods \cite{ANiklasson16,RSchade22}. In these methods the electronic degrees of freedom are either propagated dynamically using extended Lagrangian Born-Oppenheimer molecular dynamics (XL-BOMD) \cite{ANiklasson08,PSouvatzis14,ANiklasson17,ANiklasson21b} or are extrapolated from previous time steps \cite{JKolafa03,PPulay04,JMHerbert05,ANiklasson06,TDKuhne07,RSchade22}. 
In this section we will briefly present
the most recent formulations of XL-BOMD \cite{ANiklasson20,ANiklasson2011}, where we use a shadow Born-Oppenheimer potential energy surface and a preconditioned low-rank Krylov subspace approximation for the kernel in the integration of the electronic equations of motion. We will use this updated shadow potential version of XL-BOMD to improve our original graph-based QMD simulation framework \cite{ANiklasson16}. Previously, the graph-based XL-BOMD simulations sometimes could require a few SCF iterations in each time step for small-gap systems with unsteady ground-state charge solutions.
With the more recent formulation of XL-BOMD it is possible to efficiently simulate more challenging and charge sensitive systems, e.g.\ systems where the electronic HOMO-LUMO gap is opening and closing along the molecular trajectories, without any SCF iterations \cite{ANiklasson20}. Ordinarily, electronic degeneracies appearing at the chemical potential when the gap is closing can cause costly convergence problems and difficulties achieving accurate conservative forces, which may lead to unphysical molecular-dynamics trajectories. The most recent shadow potential formulation of XL-BOMD, including finite electronic temperatures and fractional occupation numbers, can avoid these problems.

\subsection{Shadow Potential Energy Surface}

The shadow potential formulation of XL-BOMD is based on a general principle that is quite simple, yet very powerful. 
Instead of calculating expensive approximate forces for the underlying exact interatomic potential, $U({\bf R})$, we design an approximate 
shadow potential, ${\cal U}({\bf R},n)$, that closely follows the exact solution, but for which we can calculate exact forces cheaply. 
This approach is usually referred to as a backward error analysis or a shadow Hamiltonian approach 
\cite{HYoshida90,CGrebogi90,SToxvaerd94,GJason00,RDEngel05,ShadowHamiltonian,SToxvaerd12,KDHammonds20}
and is often associated with the design of symplectic or geometric integration schemes 
\cite{EForest90,JPChannel90,McLachlan92,BJLeimkuhler94,JFinkelstein20}. 

In XL-BOMD the shadow potential, ${\cal U}({\bf R},n)$, depends not only on the nuclear coordinates, ${\bf R}$, but also on the electron density, $n({\bf r})$.
Thus, in contrast to regular QMD, and in the spirit of Car-Parrinello dynamics \cite{RCar85}, XL-BOMD includes the electronic degrees
of freedom, $n({\bf r})$, and its time derivative, ${\dot n}({\bf r})$, as additional extended dynamical field parameters, apart from the nuclear coordinates, ${\bf R}$, and their velocities, ${\bf \dot R}$. The equations of motion in XL-BOMD are derived in an adiabatic limit, where
the extended electronic degrees of freedom are assumed to be fast compared to the nuclear motion. This corresponds to the same basic assumption (but for a classical field) as in the Born-Oppenheimer approximation. A similar adiabatic condition can also be enforced in Car-Parrinello molecular dynamics using Lagrange multipliers, which forms the basis of the mass-zero constrained dynamics scheme by Bonella {\em et al.} \cite{SBonella20,ACoretti20,ACoretti22}.

In regular Born-Oppenheimer molecular dynamics based on density functional theory (DFT) \cite{hohen,KohnSham65,RParr89,RMDreizler90}, the potential energy surface and ground state density are given by
\begin{align}
&\rho_{\rm min}({\bf r}) = \arg \min_{\rho} \left\{E({\bf R},\rho) \left \vert \int \rho({\bf r}) d{\bf r} = N_e \right. \right\}, \label{RhoMin}\\
&U({\bf R}) = E({\bf R},\rho_{\rm min}).
\end{align}
Here $E({\bf R},\rho)$ is an energy functional that we assume also includes the ion-ion repulsive terms, and for the finite temperature generalization, including fractional occupation numbers, it is a free energy functional that also includes the electronic entropy contribution \cite{NMermin63,NMermin65,RParr89,MWeinert92,RWentzcovitch92,ANiklasson21b}. 
With the constrained minimization of $E({\bf R},\rho)$ we always mean the {\em lowest stationary solution} over all physically relevant electron densities that integrates to the desired $N_e$ number of electrons. The stationary ground state solution, $\rho_{\rm min}({\bf r})$, determines the Born-Oppenheimer potential, $U({\bf R})$, which generates the dynamics. The motion is driven by the interatomic forces, $-\nabla_{{\bf R}_I} U({\bf R})$, as given by Newton's equation of motion,
\begin{align}
    M_I {\bf \ddot R}_I =  -\nabla_{{\bf R}_I} U({\bf R}),
\end{align}
where $\{M_I\}$ are the atomic masses.

Notice that when we include the effects of finite electronic temperatures, we allow for thermally excited states as given by the fractional occupation numbers in Kohn-Sham DFT \cite{NMermin63,NMermin65,RParr89}. In principle, this goes beyond the regular ground-state Born-Oppenheimer approximation. However, here we consider also an instantaneously {\em thermally} equilibrated ground state solution a part of a generalized Born-Oppenheimer approximation.

In general, $E({\bf R},\rho)$, is a nonlinear energy functional. The constrained minimization, Eq.\ (\ref{RhoMin}), over all physically relevant electron densities, which in Kohn-Sham density functional theory 
are determined by single-particle states, leads to a system of non-linear eigenvalue equations. This non-linear problem needs 
to be solved iteratively with an SCF optimization procedure. Alternatively, the ground state can be found with an iterative direct energy minimization scheme, which in many ways resembles the iterative SCF optimization procedure. The ground-state optimization
can be quite expensive and unless the solution is well converged, the calculated forces from $U({\bf R})$
may not be sufficiently conservative, which could invalidate the results of a QMD simulation. This shortcoming can be avoided with a backward error analysis or a shadow Hamiltonian approach.

In our backward error analysis or shadow Hamiltonian approach we construct the approximate shadow potential, ${\cal U}({\bf R}, n)$,
by first approximating the energy functional, $E({\bf R},\rho)$, with an approximate $n$-dependent ``shadow'' energy functional,
\begin{align}
{\cal E}({\bf R},\rho,n) = E({\bf R},\rho) + {\cal O}\left( \vert \rho - n\vert^2 \right).
\end{align}
This can be achieved by a functional expansion (linearization) of $E({\bf R},\rho)$ around some test density, $n({\bf r})$, which we assume is an electron density that is close to the exact regular ground-state density, $\rho_{\rm min}({\bf r})$, in Eq.\ (\ref{RhoMin}). The $n$-dependent fully relaxed ground state density, $\rho_0[n]({\bf r})$, which is given from the constrained minimization of the approximate energy functional, i.e.\
\begin{align}
&\rho_0[n]({\bf r}) = \arg \min_\rho \left\{{\cal E}({\bf R},\rho,n) \left \vert \int \rho({\bf r}) d{\bf r} = N_e \right. \right\} \label{rho_0},
\end{align}
defines our $n$-dependent shadow Born-Oppenheimer potential,
\begin{align}
&{\cal U}({\bf R},n) = {\cal E}({\bf R},\rho_0[n],n) .\label{UShadow}
\end{align}
The minimization of the shadow energy functional in Eq.\ (\ref{rho_0}), which is performed with respect to the lowest stationary solution, thus defines the $n$-dependent ground state density, $\rho_0[n]({\bf r})$, and the corresponding shadow Born-Oppenheimer potential, ${\cal U}({\bf R},n)$.
The shadow energy functional, ${\cal E}({\bf R},\rho,n)$, is given by a linearization of $E({\bf R},\rho)$ in $\rho({\bf r})$ around $n({\bf r})$.  The shadow energy functional is therefore linear in $\rho({\bf r})$. The constrained minimization can then be performed directly in a single step without any SCF optimization procedure. This not only drastically reduces the cost of the optimization, but also provides forces (without any convergence problems) that are fully consistent with the shadow potential ${\cal U}({\bf R},n)$. This enables a conservative dynamics that diminishes any systematic long-term drift in the total energy in the same way as for shadow Hamiltonian integration schemes in classical molecular dynamics simulations \cite{SToxvaerd94,GJason00,RDEngel05,ShadowHamiltonian,SToxvaerd12}.

\subsection{The ground state density matrix for the shadow Born-Oppenheimer potential}

In a shadow density functional formulation for orbital-dependent Kohn-Sham theory, the minimizing electron density, $\rho_0[n]({\bf r})$, in Eq.\ (\ref{rho_0}), can be constructed indirectly through the single-particle density matrix, ${\bf D}^{\beta,{\rm ao}}[n] = \{ D_{ij}^{\beta,{\rm ao}}[n]\}$, where 
\begin{align}
&\rho_0[n]({\bf r}) = \sum_{ij} D_{ij}^{\beta,{\rm ao}}[n] \phi_i({\bf r})\phi_j({\bf r}). \label{rho_D}
\end{align}
Here we assume that $\{\phi_i({\bf r})\}$ is some real-valued atomic-orbital (indicated using the ``ao'' superscript) basis functions for the expansion of the single-particle molecular orbitals
that are used to represent the electronic density. The density matrix has the inverse temperature superscript, $\beta = 1/(k_{\rm B} T_e)$, to denote a generalization to finite electronic temperatures $(T_e \ge 0)$ with fractional occupation numbers of the molecular orbitals.  The (thermally equilibrated) ground state density matrix that is given from the minimization of the linearized energy functional in Eq.\ (\ref{rho_0}) is determined by the effective single-particle Kohn-Sham Hamiltonian, which in the basis-set representation has matrix elements,
\begin{align}
H^{\rm ao}_{ij}[{ n}] = \int \phi_i({\bf r}) \frac{\delta {\cal E}({\bf R},\rho,n)}{\delta {\rho({\bf r})}} \phi_j({\bf r}) d{\bf r}.
\end{align}
In our generalized finite temperature formulation
of Kohn-Sham density functional theory \cite{NMermin63,NMermin65,RParr89,ANiklasson21b} (or in finite temperature Hartree-Fock theory \cite{Roothaan,McWeenyHF,NMermin63})
the $n$-dependent, thermally relaxed ground state density matrix of the linearized energy functional in Eq.\ (\ref{rho_0}) is given by 
\begin{align}
&{\bf D}^\beta[n] = \left[e^{\beta ({\bf H}[n]-\mu {\bf I})}+{\bf I} \right]^{-1}.\label{Dmin_n}
\end{align}
We here assume that ${\bf H}[n]$ and ${\bf D}^\beta[n]$ are in an orthogonalized matrix representation, where ${\bf H}[n] = {\bf Z}^T{\bf H}^{\rm ao}[n]{\bf Z}$ and ${\bf D}^{\beta,{\rm ao}}[n] = {\bf Z}{\bf D}^{\beta}[n] {\bf Z}^T$,
for some inverse factorization of the overlap matrix, $S_{ij} = \int \phi_i({\bf r})\phi_j({\bf r})d{\bf r}$, such that
${\bf Z}^T{\bf S}{\bf Z} = {\bf I}$. The eigenvalues of ${\bf D}^\beta[n]$ correspond to the fractional occupation numbers of the electronic states.
Because ${\bf H}[n]$ only depends on $n({\bf r})$ and is independent of $\rho_0[n]({\bf r})$ we don't need to solve
for ${\bf D}^\beta[n]$ and $\rho_0[n]({\bf r})$ in a self-consistent manner as 
for the constrained minimization of $E({\bf R},\rho)$.
The ground state electron density, $\rho_0[n]({\bf r})$, can therefore be constructed directly without relying on an iterative self-consistent field optimization procedure. This is of particular importance in QMD simulations using linear-scaling electronic structure theory, where a numercial thresholding or other approximations can cause convergence problems \cite{MCawkwell12,MArita14,TOtsuka16,ANiklasson16,THirakawa17}. The same benefit has been demonstrated for QMD simulations with AI-accelerating Tensor cores, where a low  floating-point precision is used for the electronic structure calculations \cite{JFinkelstein21,JFinkelstein21B}. The graph-based QMD simulations will benefit in the same way.

\subsection{Equations of motion}

The optimized $n$-dependent ground state density, $\rho_0[n]({\bf r})$, calculated from the density matrix in Eq.\ (\ref{Dmin_n}) above, defines the shadow Born-Oppenheimer potential, ${\cal U}({\bf R},n)$, in Eq.\ (\ref{UShadow}). The error in the shadow potential can be estimated from the size of a residual function, $f [n]({\bf r}) =\rho_0[n]({\bf r})-n({\bf r})$, where
\begin{equation}
   {\cal U}({\bf R},n) - U({\bf R}) = {\cal O}(\vert \rho_0[n]-n\vert^2). \label{UErr}
\end{equation}
The deviation from the exact regular Born-Oppenheimer potential thus depends on how close $n({\bf r})$ is to the ground state density, $\rho_0[n]({\bf r})$, or $\rho_{\rm min}({\bf r})$ \footnote{When $n=\rho_0[n]$ we have an exact self-consistent solution and then $\rho_0[n] = \rho_{\rm min}$.}. To keep $n({\bf r})$ close to ground state as the positions of the atoms evolve during a molecular dynamics simulation, we propagate $n({\bf r})$ as a dynamical field variable that is driven by an extended harmonic oscillator that is centered on the ground state density. This additional dynamics can be formulated with an extended Lagrangian, using our shadow Born-Oppenheimer potential, and where $n({\bf r})$ and its time derivative are included as extended electronic degrees of freedom in addition to the atomic positions, ${\bf R} = \{{\bf R}_I\}$, and their velocities, ${\bf \dot R} = \{{\bf \dot R}_I\}$ \cite{ANiklasson08,ANiklasson17,ANiklasson21b}. The equations of motion are given from the Euler-Lagrange equations that are derived in an adiabatic limit, where
we assume that the extended electronic degrees of freedom are fast compared to the fastest nuclear motion. This is analogous and consistent with the regular Born-Oppenheimer approximation, which was our starting point in the definition of the Born-Oppenheimer potential. 
In this adiabatic limit the equations of motions of our shadow Born-Oppenheimer molecular dynamics are given by
\begin{align}
&M_I {\ddot {\bf R}}_I = \left. \nabla_{{\bf R}_I}{\cal U}({\bf R},n) \right \vert_n, \label{NEQ}\\
&{\ddot n}({\bf r}) = - \omega \int K({\bf r,r'}) \left(\rho_0[n]({\bf r'}) - n({\bf r'})\right)d{\bf r'} \label{HW}.
\end{align}
where $\{M_I\}$ are the atomic masses.
The first equation, Eq.\ (\ref{NEQ}), is the regular Newton's equation of motion for the nuclear degrees of freedom with the forces calculated from the gradient of the shadow potential, while the density, $n({\bf r})$, is kept constant.
The second equation, Eq.\ (\ref{HW}), is a harmonic oscillator equation for the electron density, $n({\bf r})$, that evolves in a harmonic well centered on $\rho_0[n]({\bf r})$. In the harmonic oscillator equation $K({\bf r,r'})$ is a kernel defined by
\begin{align}\label{KDef}
\int K({\bf r,r'})\frac{\delta \left(\rho_0[n]({\bf r''})- n({\bf r''})\right)}{\delta n({\bf r'})} d{\bf r'} = \delta({\bf r-r''}).
\end{align}
This kernel appears similar to a preconditioner, which makes $n({\bf r})$ oscillate around a close approximation to the fully self-consistent electron density of the regular Born-Oppenheimer potential, $\rho_{\rm min}({\bf r})$,
and therefore also to $\rho_0[n]({\bf r})$.
In this way $n({\bf r})$ follows the ground state and the error in the shadow potential, Eq.\ (\ref{UErr}), stays small.
Typically, the kernel is approximated using a scaled delta function, i.e.\ $K({\bf r,r'}) = -c \delta({\bf r-r'}), ~~ c \in [0,1]$.
However, for sensitive small-gap systems with unsteady charge solutions, this is often not sufficient and a more accurate, preconditioned, low-rank Krylov subspace approximation is needed \cite{ANiklasson17}. This will be discussed in more detail below and it is the key motivation behind our introduction of a graph-based quantum perturbation theory.

The integration of the equations of motion, in Eq.\ (\ref{NEQ}) and Eq.\ (\ref{HW}), can be performed with a modified Verlet integration scheme \cite{ANiklasson09,PSteneteg10,GZheng11}. To initialize the electronic degrees of freedom we set $n({\bf r}) = \rho_{\rm min}({\bf r})$; this requires a full regular SCF optimization of the ground state density as in Eq.\ (\ref{RhoMin}), but only for the first time step.  The accuracy in the sampling of the potential energy can then be shown to depend on the integration time step, $\delta t$, to fourth order \cite{ANiklasson17}, i.e. 
\begin{align}
&\vert {\cal U}({\bf R},n) - U({\bf R})\vert = {\cal O}\left( \vert \rho_0[n] - n\vert^2 \right) \propto \delta t^4.
\end{align}
The sampling error of the potential surface is thus small compared to the local truncation error in the total energy for the Verlet integration scheme, where the accuracy in the total energy scales as ${\cal O}(\delta t^2$). In general, we can therefore use the same size of the integration time step as in regular direct Born-Oppenheimer molecular dynamics simulations \cite{ANiklasson21b}.

\subsection{Approximating the kernel}

The kernel $K({\bf r,r'})$ in Eq.\ (\ref{HW}) is defined by Eq.\ (\ref{KDef}) as the inverse Jacobian of the residual
function, 
\begin{equation}
    f[n]({\bf r}) = \rho_0[n]({\bf r})- n({\bf r}).
\end{equation}
The calculation of the kernel can be a demanding task. However, we only need to approximate how the kernel acts on the residual function \cite{ANiklasson20}. This allows for a significant simplification.

To facilitate the presentation of how the kernel, $K({\bf r,r'})$, can be approximated we will use matrix-vector notation. 

In a matrix-vector notation, the kernel is given by
\begin{align}
{\bf K} = {\bf J}^{-1}, ~~ J_{ij} = \frac{\partial f_i({\bf n})}{\partial n_j}, ~~{\bf f} = \left({\boldsymbol \rho}_{ 0}[{\bf n}] - {\bf n} \right), \label{Jacobian}
\end{align}
where ${\bf K}, {\bf J} \in {\boldmath R}^{N \times N}$, and ${\bf f},{\bf n}, {\boldsymbol \rho}_{ 0}[{\bf n}] \in {\boldmath R}^{N}$.
We can then rewrite the electronic equations of motion in Eq.\ (\ref{HW}) in the equivalent form,
\begin{align}
&{\bf \ddot n} = - \omega^2\left({\bf K}_0{\bf J}\right)^{-1} {\bf K}_0\left({\boldsymbol \rho}_{\rm 0}[{\bf n}] - {\bf n} \right). \label{EEOM}
\end{align}
We have here introduced an approximate inverse to the Jacobian, ${\bf K}_0 \approx {\bf J}^{-1}$.
This means that $({\bf K}_0{\bf J})^{-1} \approx {\bf I}$, whose action on the preconditioned residual vector, ${\bf K}_0{\bf f}$, should be possible
to represent accurately using a low-rank approximation. In the exact case we can represent $({\bf K}_0{\bf J})^{-1}$ by
\begin{align}
&({\bf K}_0{\bf J})^{-1} = \sum_{ij}^N {\bf v}_i M_{ij} {\bf f}_{{\bf v}_j}^T, \label{PreKernel}
\end{align}
where
\begin{align}
& {\bf f}_{{\bf v}_j} \equiv {\bf K}_0\left. \frac{\partial {\bf f}({\bf n} + \lambda {\bf v}_j)}{\partial \lambda }\right \vert_{\lambda = 0} = {\bf K}_0{\bf J} {\bf v}_j, \label{dfdv}
\end{align}
for some complete set of vectors, $\{{\bf v}_j\}$.
The tensor elements, $\{M_{ij}\}$, are given by
\begin{align}
{\bf M} = {\bf O}^{-1}, ~~O_{ij} = {\bf f}_{{\bf v}_i}^T {\bf f}_{{\bf v}_j}.
\end{align}
The expression for the preconditioned kernel in Eq.\ (\ref{PreKernel}) is based on a generalization of the Jacobian, ${\bf J}$, which is defined by an arbitrary set of directional derivatives \cite{ANiklasson20} instead of the partial derivatives as in Eq.\ (\ref{Jacobian}). 

The directional derivatives in Eq.\ (\ref{dfdv}) along ${\bf v}_j$ can be calculated using time-independent quantum perturbation theory. The charge vector, ${\bf v}_j$, generates a linear perturbation, ${\bf H}_1$, in the potential of the unperturbed Kohn-Sham Hamiltonian, ${\bf H}_0$, i.e.\ we get a linear first-order perturbation in the Hamiltonian, where ${\bf H}(\lambda) = {\bf H}_0 + \lambda {\bf H}_1$. The first-order response in the charge density is then given from the first-order perturbation in the wavefunctions or the density matrix. How we can perform such response calculations on a partitioned graph, including fractional occupation numbers, will be presented in the next section below and is one of the main objectives of this paper.

A low-rank approximation of $({\bf K}_0{\bf J})^{-1}$ in Eq.\ (\ref{PreKernel}), which is acting on ${\bf K}_0{\bf f}({\bf n})$ in Eq.\ (\ref{EEOM}), can be constructed with a reduced set of well-chosen vectors $\{{\bf v}_j\}_{j=1}^m$ $(m < N)$.
We can chose these from an orthogonalized preconditioned Krylov subspace,
\begin{align}
&\{{\bf v}_j\} = {\rm span}^\perp \left\{{\bf K}_0{\bf f}({\bf n}), ({\bf K}_0{\bf J})^1{\bf K}_0{\bf f}({\bf n}), \right. \\
&~~~~~~~~~~\left. ({\bf K}_0{\bf J})^2{\bf K}_0{\bf f}({\bf n}), ({\bf K}_0{\bf J})^3{\bf K}_0{\bf f}({\bf n}), \ldots   \right\}.
\end{align}
With this preconditioned Krylov subspace approximation, which is described in more detail in Sec.\ \ref{KrylovExp} below, we can rewrite the electronic equations of motion with a low-rank expression as
\begin{align}
{\bf \ddot n} \approx - \omega^2 \left(\sum_{ij}^{m < N} {\bf v}_i M_{ij} {\bf f}_{{\bf v}_j}^T \right){\bf K}_0\left({\boldsymbol \rho}_{\rm 0}[{\bf n}] - {\bf n} \right). \label{LowRankEEOM}
\end{align}
The main cost of this preconditioned kernel approximation, which was introduced in Ref.\ \cite{ANiklasson20} for the integration of the extended electronic degrees of freedom in XL-BOMD, comes from the calculation of the response vectors, ${\bf f}_{{\bf v}_j}$. These can be calculated using canonical quantum perturbation theory that accounts for fractional occupation numbers \cite{ANiklasson15,YNishimoto17}. In our graph-based approach we need to modify this canonical perturbation theory to be applicable to calculations on a graph. This is particularly important for sensitive low-gap systems that may have unsteady charge solutions.
However, for QMD simulations of non-reactive systems, we can often avoid
any low-rank updates of ${\bf f}_{{\bf v}_j}$. Instead, a simple fixed preconditoner, ${\bf K}_0$, alone (with $m=0$) is sufficiently accurate. Typically, for molecular systems with a significant HOMO-LUMO gap even a scaled delta function, where ${\bf K} \approx {\bf K}_0 = -c{\bf I}$ with $c \in [0,1]$, provides an accurate approximation. Alternatively, a preconditioner can be constructed from the regularized solution of some approximate system, for example, the molecular system at the initial time step. Even if the calculation of the preconditioner is expensive, if it is reused over many (e.g. thousands) time steps, the total overhead may become only a small fraction of the total computational cost, leading to a net increase in efficiency.
 
\section{Canonical quantum perturbation theory on a graph}\label{GraphPRT}

The time-independent canonical quantum response calculations required for the directional derivatives in Eq.\ (\ref{dfdv}), which are used in the preconditioned kernel approximation in Eq.\ (\ref{LowRankEEOM}), can be formulated in terms of  density matrix perturbation
theory \cite{RMcWeeny62,ANiklasson04,VWeber04,ANiklasson15,LTruflandier20,ANiklasson20}. Density matrix perturbation theory for fractional occupation numbers can be constructed from matrix function expansions \cite{ANiklasson15}. We can therefore design a graph-based canonical quantum perturbation theory if we use the one-to-one mapping between the matrix function expansion on a graph and the collection of the core parts from the dense matrix functions of the principal submatrices that are determined by the partitioned subgraphs, i.e.\ as in Fig.\ \ref{Fig_1} or in Eq.\ (\ref{dCollect}). First we show how the time-independent linear response in the density matrix can be calculated on a partitioned graph and then we look at the response in observables, and in particular the linear response in the electron density.

\subsection{Linear response in the density matrix} 

The unperturbed effective single-particle density matrix for
a system at finite electronic temperature is given, as in Eq.\ (\ref{Dmin_n}), by 
\begin{align}
{\bf D}_0^\beta &=\Theta_\beta \left(\mu {\bf I} - {\bf H}_0\right) \equiv \left[e^{\beta ({\bf H}_0-\mu {\bf I})}+{\bf I} \right]^{-1}.
\end{align}
Here ${\bf H}_0$ is the unperturbed Hamiltonian, which we, once again, assume is in an orthogonal
representation. The chemical potential, $\mu$, is determined by the condition that the trace of the density matrix has a given occupation number, i.e.\
\begin{align}
&{\rm Tr}\left[ {\bf D}_0^\beta\right] = N_{\rm occ} ~\Rightarrow \mu.
\end{align}
If we now introduce a perturbation to the Hamiltonian,
\begin{align}
&{\bf H} \equiv {\bf H}(\lambda) = {\bf H}_0 + \lambda {\bf H}_1,
\end{align}
that is linear in the perturbation parameter, $\lambda$, then the corresponding linear response
in the density matrix is
\begin{align}
{\bf D}_1^\beta &=  \left. \frac{\partial \Theta_\beta \left(\mu {\bf I} - {\bf H}(\lambda)\right) }{\partial \lambda} \right \vert_{\lambda = 0} \label{D1_a}\\
&~~~~~~~~~~~~~~ + \frac{\partial \Theta_\beta \left(\mu {\bf I} - {\bf H}(\lambda)\right) }{\partial \mu}\left. \frac{\partial \mu}{\partial \lambda} \right \vert_{\lambda = 0}.\label{D1_b}
\end{align}
The separate response terms of ${\bf D}_1^\beta$ can be calculated using first-order canonical density matrix perturbation theory \cite{ANiklasson15,YNishimoto17,ANiklasson20}. The first partial derivative term in Eq.\ (\ref{D1_a}) corresponds to a grand-canonical response with a fixed chemical potential and can be calculated, for example, as the ${\bf X}^{(1)}$ term in Alg.\ 2 in Ref.\ \cite{ANiklasson20b}. The second term in Eq.\ (\ref{D1_b}) is the response of the density matrix with respect to
the chemical potential, $\mu$, which is given by
\begin{align}
& \left. \frac{\partial \Theta_\beta \left(\mu {\bf I} - {\bf H}(\lambda)\right) }{\partial \mu}\right \vert_{\lambda = 0} = \beta {\bf D}_0^\beta \left({\bf I}-{\bf D}_0^\beta \right).
\end{align}
To calculate the last factor, $\mu_1 \equiv {\partial \mu}\big / {\partial \lambda}$, which is the response in the chemical potential, we can use the condition that the perturbation does not change the number of states, i.e.\
\begin{align}
&{\rm Tr} \left[ {\bf D}_1^\beta \right] = 0.
\end{align}
This condition gives us the response term,
\begin{align}
& \mu_1 = - \frac{\rm Tr\left[ \partial \Theta_\beta \left(\mu {\bf I} - {\bf H}\right)\big / \partial \lambda \right]}
{\rm Tr\left[ \partial \Theta_\beta \left(\mu {\bf I} - {\bf H}\right) \big / \partial \mu \right]}.
\end{align}
For convenience we have dropped the notation that the response terms are calculated in the limit of $\lambda = 0$.

To perform the corresponding canonical quantum response calculations on a graph, we can break up the data dependency
graph, ${\bf G}_\tau$, into subgraphs from which we can extract the principal submatrices, $\{{\bf h}_0^{(i)}\}$
and $\{ {\bf h}_1^{(i)} \}$ from ${\bf H}_0$ and its perturbation ${\bf H}_1$, respectively. 
The first-order response in the density matrix, ${\bf D}_1^\beta\big \vert_{G_\tau}$, calculated on the data connectivity graph, ${\bf G}_\tau$, is then given by the collected linear response terms over the core parts of the separate susbsystems, i.e.\
\begin{align}
{\bf D}_1^\beta\big \vert_{G_\tau} &= \left\{ {\bf d}_{1,\lambda}^{\beta,(i)} \Big \vert_{\rm c}\right\}_{\rm collect} + \mu_1 \left\{ {\bf d}_{1,\mu}^{\beta,(i)} \Big\vert_{\rm c}\right\}_{\rm collect}. \label{D1}
\end{align}
Here
\begin{align}
&{\bf d}_{1,\lambda}^{\beta,(i)} = \left. \frac{\partial \Theta_\beta \left(\mu {\bf I}^{(i)} - {\bf h}^{(i)}_0 - \lambda  {\bf h}^{(i)}_1\right) }{\partial \lambda} \right \vert_{\lambda = 0}, \label{ddlambda}
\end{align}
and
\begin{align}
&{\bf d}_{1,\mu}^{\beta,(i)} = \beta {\bf d}^{\beta,(i)}_{0}\left({\bf I}^{(i)}-{\bf d}^{\beta,(i)}_{0}\right). \label{dddmu}
\end{align}
These are the density matrix response terms for the local subsystems. The response in the chemical
potential that keeps the total number of electrons unchanged is then given by
\begin{align}
&\mu_1 = - \frac{\sum_i {\rm Tr}\left[{\bf d}_{1,\lambda}^{\beta,(i)} \Big \vert_{\rm c} \right]}
{\sum_i {\rm Tr}\left[{\bf d}_{1,\mu}^{\beta,(i)} \Big\vert_{\rm c} \right]}, \label{dmu_1}
\end{align}
where the traces are taken over the core part of each subsystem. 
The density matrix response calculations of $\{{\bf d}_{1,\lambda}^{\beta,(i)}\}$ and $\{ {\bf d}_{1,\mu}^{\beta,(i)} \}$ on the separate subgraphs can be performed in the molecular-orbital representation, e.g.\ as in Alg.\ 2 in Ref.\ \cite{ANiklasson20b}. This is particularly efficient when we need multiple response evaluations for the same structure and when we need to consider fractional occupation numbers. 

\subsection{Response in observables}

With the linear response in the density matrix we can calculate the response of any time-independent observable, $a$, corresponding to some operator ${\bf \hat A}$. In a bra-ket representation, where $A_{ij} = \langle \phi_i\vert{\bf \hat A}\vert \phi_j \rangle$, the response is given by
\begin{equation}
    a(\lambda) = Tr[{\bf A}{\bf D}_0^\beta] + \lambda Tr[{\bf A}{\bf D}_1^\beta].
\end{equation}
In this way we can calculate, for example, the polarizability from the response of an electric field, if ${\bf \hat A}$ is the dipole operator \cite{vweber05,ANiklasson15}. 

From Eq.\ (\ref{rho_D}) we find that the linear response in the ground state electron density, $\rho_0[n]({\bf r})$, is given by
\begin{equation}
    \rho_\lambda[n]({\bf r}) = \rho_0[n]({\bf r}) + \lambda \rho_1[n]({\bf r}),
\end{equation}
where
\begin{equation}  
     \rho_1[n]({\bf r}) = \sum_{ij} { D}_{1,ij}^{\beta,{\rm ao}}[n] \phi_i({\bf r}) \phi_j({\bf r}) \label{drho_1}.
\end{equation}
Here ${\bf D}_1^{\beta,{\rm ao}}[n]$ is the non-orthogonal atomic-orbital representation of the linear response in the density matrix, i.e.\ ${\bf D}_1^{\beta,{\rm ao}}[n] = {\bf Z} {\bf D}_1^{\beta}[n] {\bf Z}^T$, where ${\bf Z}$ is the inverse factorization of the overlap matrix, ${\bf S}$, i.e.\ where ${\bf Z}^T{\bf S}{\bf Z} = {\bf I}$. 

In general, we never need to construct the full collected density matrix response matrix, ${\bf D}^{\beta}_1 \big \vert_{G_\tau}$ from the core parts of the separate subgraph density matrix responses, $\{{\bf d}_{1,\lambda}^{\beta,(i)}\}$ and $\{ {\bf d}_{1,\mu}^{\beta,(i)} \}$. Instead, response properties, such as the response in the electron density or the response of the partial charges, can be calculated first for the separate subsystems and then collected. In this way the amount of data transfer can be reduced when the response calculations are performed on a distributed parallel platform. 

The data dependency graph, ${\bf G}_\tau$, that is needed for the graph-based quantum response calculations can be determined, not only from an estimate of the globally thresholded expansions of the ground state density matrix, ${\bf D}_0^\beta$, and the Kohn-Sham Hamiltonian, ${\bf H_0}$, but also from the response density matrix, ${\bf D}_1^\beta$. The matrix ${\bf D}_1^\beta$ is often less sparse than ${\bf D}_0^\beta$ \cite{vweber05}; however, in practice, we have found that it is sufficient to estimate ${\bf G}_\tau$ from the same graph structures of a thresholded ${\bf D}_0^\beta$ and ${\bf H_0}$, as discussed in Sec.\ \ref{Estimate_G}.

Our graph-based canonical quantum perturbation theory was motivated by the need to calculate the response in the electron density, as in Eq.\ (\ref{dfdv}), which is needed to approximate the kernel in our shadow potential formulation of XL-BOMD \cite{ANiklasson20}. 
Graph-based quantum perturbation theory is a very useful and general tool in its own right, however. In addition, beyond the present application, it should also be possible to use the theory to reduce the complexity of calculations of a broad range of time-independent response properties of large extended systems, including properties such as the magnetic susceptibility, phonon modes, electric polarizabilities, the Born-effective charge, and Raman spectra \cite{SBaroni01,ANiklasson03,VWeber04,COchsenfeld04}.

\section{Graph-based shadow Born-Oppenheimer molecular dynamics}

The equations of motion that determine our shadow Born-Oppenheimer molecular dynamics, in Eqs.\ (\ref{NEQ}) and (\ref{HW}), can be rewritten in an equivalent matrix-vector notation as
\begin{align}
&M_I {\ddot {\bf R}}_I = \left. \nabla_{{\bf R}_I}{\cal U}({\bf R},{\bf n}) \right \vert_{\bf n}, \label{NEQ_2}\\
&{\bf \ddot n} = - \omega^2\left({\bf K}_0{\bf J}\right)^{-1} {\bf K}_0\left({\boldsymbol \rho}_{\rm 0}[{\bf n}] - {\bf n} \right), \label{EEQOM}
\end{align}
where we have included the preconditioner, ${\bf K_0}\approx {\bf J}^{-1}$, in the electronic equations of motion. 

Later, in our demonstration of the graph-based shadow Born-Oppenheimer molecular dynamics simulations, we will use 
semi-empirical SCC-DFTB theory, which is naturally described using the vector notation.  
The theory is general, however, and can be applied to electronic structure methods with different representations of the electronic structure, perhaps with some significant modifications.

This section mirrors and partially overlaps with the response theory discussed in the previous section \ref{GraphPRT}. However, the analysis in this section complements the previous discussion. Instead of a general presentation of a graph-based canonical density matrix perturbation theory we now focus on the specific techniques required for the graph-based shadow Born-Oppenheimer molecular dynamics simulations. First we look at the construction of a preconditioner and then how the low-rank preconditioned Krylov subspace approximation can be performed on a graph for the integration of the electronic equation of motion.

\subsection{Preconditioner for a partitioned graph}

The effictiveness of the Krylov subspace approximation of $\left({\bf K}_0{\bf J}\right)^{-1}$ acting on the preconditioned residual function,
${\bf K}_0\left({\boldsymbol \rho}_{\rm 0}[{\bf n}] - {\bf n} \right)$, in Eq.\ (\ref{EEQOM}) depends on how close the preconditioner, ${\bf K}_0$, is to the exact solution, ${\bf J}^{-1}$. 
It is possible construct ${\bf K}_0$ by calculating a regularized kernel for an approximate test system, for example, the full composite system at the initial time step of a QMD simulation or for some similar equilibrated structure. 
The principal submatrices of ${\bf K}_0$ can then be extracted on-the-fly during an MD simulation as preconditioners for the partitioned subgraphs. Such a preconditioner could be reused, possibly over thousands of time steps. Without a parallel implementation, the computational cost currently limits this approach to systems with $\lesssim$O($10^4$) atoms. Fortunately our 
graph-based methods enable a modified parallel approach where the preconditioner for the whole system is decomposed into a set of smaller separate preconditioners, one for each subgraph. This approach must account for the possibility that the halos surrounding the core regions of the subgraphs can change in each time step, which increases the complexity. It is also necessary to take care that application of the preconditioner conserves the total charge. We now describe such an approach. 

The kernel ${\bf K}$ is given from the inverse Jacobian, ${\bf J}^{-1}$, of the residual vector function,
\begin{align}
&{\bf f}({\bf n}) = {\boldsymbol \rho}_0[{\bf n}] - {\bf n}.
\end{align}
The Jacobian matrix elements are given by
\begin{align}
&{J}_{ij} = \frac{\partial {f_i}({\bf n})}{\partial {n_j} } + \frac{\partial {f_i}({\bf n})}{\partial \mu }\frac{\partial \mu}{\partial {n_j}} \label{Jacob}\\
& = \frac{\partial {{\rho}_0}_i[{\bf n}]}{\partial {n_j} } + \frac{\partial {{\rho}_0}_i[{\bf n}]}{\partial \mu }\frac{\partial \mu}{\partial {n_j}} - \delta_{ij}, \label{TwoDer}
\end{align}
where the response of the chemical potential, $\mu$, has been included explicitly.

The ${\bf n}$-dependent ground-state charge vector, ${\boldsymbol \rho}_0[{\bf n}]$, which is given from the constrained minimization in Eq.\ (\ref{rho_0}),
is determined indirectly from the density matrix as in Eq.\ (\ref{rho_D}). The ground state electron density can thus be seen as a function of the density matrix, where
\begin{align}
&{\boldsymbol \rho}_0[{\bf n}] \equiv {\boldsymbol \rho}_0({\bf D}^\beta_0[{\bf n}]).
\end{align}
With the graph-based scheme the ground state electron density therefore can be determined by collecting the charges of the core parts of the subgraphs, $\{{\boldsymbol \rho}_0^{(k)}[{\bf n}]\vert_c\}$. These depend on the submatrices of the density matrix corresponding to the partitioned subgraphs, $\{{\bf d}^{\beta,(k)}_0[{
\bf n}]\}$. The composite charges of the full system therefore can be assembled as
\begin{align}
&{\boldsymbol \rho}_0[{\bf n}] \Big \vert_{G_\tau} = \left\{  {\boldsymbol \rho}_0^{(k)}( {\bf d}^{\beta,(k)}_0[{
\bf n}]) \Big \vert_{\rm c}  \right\}_{\rm collect}.\label{assemble_rho0}
\end{align}
Equation~(\ref{assemble_rho0}) allows us to determine the partial derivatives necessary to calculate the Jacobian as in Eqs.\ (\ref{Jacob}) and (\ref{TwoDer}), using partial derivatives with respect to individual point charges. 
The first derivative in Eq.\ (\ref{TwoDer}) of the ground state charge is
\begin{align}
&  \frac{\partial {{\rho}_0}_i[{\bf n}]}{\partial {n_j} } =  \left\{  \rho_{0i}^{(k)}( {\bf d}^{\beta,(k)}_{1,\delta_j}[{
\bf n}]) \Big \vert_{\rm c}  \right\}_{\rm collect},
\end{align}
where
\begin{align}
&{\bf d}_{1,\delta_j}^{\beta,(k)}[{\bf n}] = \frac{\partial \Theta_\beta \left(\mu {\bf I}^{(k)} - {\bf h}^{(k)}_0[{\bf n}]- \lambda_j {\bf h}_{1}^{(k)} \right) }{\partial \lambda_j}\Big\vert_{\lambda_j = 0}.
\end{align}
Here $\{ {\bf h}_{1}^{(k)}\}$ are the principal submatrices (in an orthogonal representation) of the  potential matrix, ${\bf H}_1[\delta_j]$, from a unit delta charge, $\delta_j$, placed at $n_j$, and with all other set to zero, i.e.\ $n_{k} = 0, \forall k \ne j$. This approach requires a full (non-local) Coulomb summation to construct ${\bf H}_1[\delta_j]$, one for each charge component, $n_j$, of the system. This step could be expensive, but it can be parallelized easily and the Coulomb summations can be performed using a fast Fourier transform \cite{JCooley65}, the Ewald summation method, or the particle mesh Ewald algorithm \cite{TDarden93} \footnote{Notice, this approach needs to be adapted to work with other representations or the electronic degrees of freedom besides some coarse grained charge density, e.g.\ of partial atomic charges.}.
The second term in Eq.\ (\ref{TwoDer}) contains the derivative of the ground state charge for each component $i$ with respect to the chemical potential, $\mu$, which can be collected from the submatrices,
\begin{align}
&  \left. \frac{\partial {{\rho}_0}_i[{\bf n}]}{\partial {\mu} }\right \vert_{G_\tau} =  \left\{  \rho_{0i}^{(k)}( {\bf d}^{\beta,(k)}_{1,\mu}[{
\bf n}]) \Big \vert_{\rm c}  \right\}_{\rm collect}.
\end{align}
Here $\{ {\bf d}^{\beta,(k)}_{1,\mu}\}$ are given as in Eq.\ (\ref{dddmu}).
The last remaining response part of Eq.\ (\ref{TwoDer}) is the derivative of the chemical potential with respect to the charge, $\mu_{1,j} \equiv \partial \mu\big /\partial n_j$, which is equal for all subsystems, i.e.\ $\mu_{1,j}^{(k)} = \mu_{1,j}$. This shared response in the chemical potential can be determined from the condition that we have no net changes in the total occupation of the collected response density matrices for the full system, i.e.\ in the same way as in Eq.\ (\ref{dmu_1}), which gives us
\begin{align}
&\sum_{i \in {\rm c}} \left(\rho_{0i}^{(k)}( {\bf d}^{\beta,(k)}_{1,{\delta}_j}[{
\bf n}]) \Big \vert_{\rm c}   + \rho_{0i}^{(k)}( {\bf d}^{\beta,(k)}_{1,\mu}[{\bf n}]) \Big \vert_{\rm c}  \mu_{1,j} \right)  = 0,
\end{align}
from which each $\mu_{1,j}$ can be determined.

With the separate derivatives in Eq.\ (\ref{TwoDer}) we can now construct the Jacobian and through its inverse the kernel, ${\bf K} = {\bf J}^{-1}$, of the full composite system. Because calculating this preconditioner could be quite expensive and the parallelization is not ideally suited for our graph-based methods, we adopt an alternative approach using a set of small fixed preconditioners, one for each subgraph partition. Such an approach lowers the cost of calculating the preconditioning and reduces the necessary data transfer and/or the memory requirements. The approach follows from noting that the Jacobian, $J_{ij}$, of ${\bf f}({\bf n})$ in Eq.\ (\ref{Jacob}), can be represented by a collection of separately calculated submatrix Jacobians,
\begin{align}
    J_{ij} \big \vert_{G_\tau} = \left\{ j_{ij}^{(k)} \right\}_{\rm collect},
\end{align}
where
\begin{align}
& j^{(k)}_{ij} = \left(\rho_{0i}^{(k)}( {\bf d}^{\beta,(k)}_{1,{n}_j}[{ 
\bf n}]) \Big \vert_{\rm c}  + \right. \\
& ~ \left. + \rho_{0i}^{(k)}( {\bf d}^{\beta,(k)}_{1,\mu}[{
\bf n}]) \Big \vert_{\rm c}  \mu_{1,j}^{(k)} - \delta^{(k)}_{ij} \big \vert_{\rm c}\right). 
\end{align}
The core part of each of these Jacobians, ${\bf j}^{(k)}_{\rm c} \in R^{m \times m}$, can be inverted, yielding  trial kernels that could be used as preconditioners. To guarantee charge neutrality when the preconditioner acts on the residual function, however, we need first to adjust the response for each core part of the Jacobians such that response of each column of ${\bf j}^{(k)}_{\rm c}$ sums up to $-1$.  
The adjustment is achieved by shifting the response in the chemical potential $\mu_{1,j}^{(k)}$ such that $\sum_{i \in c} j^{(k)}_{ij} = -1$, for each column response summed only over the core parts. After this charge-response adjustment the approximate kernels for the core parts of each subgraph are given by,
\begin{align}
& {\bf k}^{(k)}_0 =  \left( {\bf j}^{(k)}_{\rm c} - \alpha{\bf I}^{(k)}\right)^{-1}, ~ \alpha \in [0,1],
\end{align}
where we have included a constant, $\alpha \ge 0$, that introduces a regularization of the solution.
The collection of these separate approximate inverse Jacobians corresponds to a block-diagonal preconditioner for the composite full system,
\begin{align}
{\bf K}_0  = \left\{ {\bf k}^{(k)}_0 \right\}_{\rm collect}. \label{PreCond}
\end{align}
There are of course other alternatives for how a preconditioner can be constructed, but we have found that this approach is particularly simple and efficient. The set of separate preconditioners, $\{{\bf k}^{(k)}_0\}$, can be distributed and used for each core part of our graph partitioning and each ${\bf k}^{(k)}_0$ does not depend other subgraphs or on the fluctuating size of the halos. 

\subsection{Preconditioned Krylov Subspace Approximation for the electronic equations of motion}\label{KrylovExp}

The block-diagonal preconditioner, ${\bf K}_0$ in Eq.\ (\ref{PreCond}), can be used to accelerate the Krylov subspace approximation of the kernel acting on the residual function in the equations of motion for the electronic degrees of freedom in Eq.\ (\ref{EEQOM}), which then can be approximated as in Eq.\ (\ref{LowRankEEOM}). 
To achieve this we need to construct the orthogonalized preconditioned Krylov subspace expansion of vectors $\{{\bf v}_i\}$ and $\{{\bf f}_{{\bf v}_i}\}$. With the graph-based approach to quantum response calculations, it is possible to perform the most demanding tasks of this Krylov subspace expansion in parallel on the separate subsystems. 
To explain how this is achieved we will show how the first vectors, ${\bf v}_1$, ${\bf f}_{{\bf v}_1}$ and ${\bf v}_2$, are constructed. We will then present the general algorithm. 

The first vector, ${\bf v}_1$, of the Krylov subspace is chosen as the normalized preconditioned residual function, i.e.\
\begin{align}
   {\bf v}_1' &= {\bf K}_0\left({\boldsymbol \rho}_{\rm 0}[{\bf n}] - {\bf n} \right) \label{v1_prim} \\
   &= \left\{{\bf k}_0^{(k)}\left({\boldsymbol \rho}_0^{(k)}[{\bf n}]\big \vert_{\rm c} - {\bf n}^{(k)}\big \vert_{\rm c}\right)\right\}_{\rm collect}\\
   & = \left\{{\bf v}_i^{(k)}\big \vert_{\rm c}\right\}_{\rm collect}\\
   {\bf v}_1 &= {\bf v}_1'\big /\|{\bf v}_1'\|. \label{v1}
\end{align}
Because the preconditoner, ${\bf K}_0$, is a block-diagonal matrix, the initial preconditioning can easily be performed on the core parts of the separate subsystems. Also the square of the vector norm, $\|{\bf v}'_1\|^2$, can be collected from the squares of the vector norms of the core parts of the separate subsystems. Once ${\bf v}_1$ is chosen, we need to calculate the directional derivative, ${\bf f}_{{\bf v}_1}$, from the preconditioner acting on the directional derivative of the residual function,
\begin{align}
&{\bf f}({\bf n}) = \left({\boldsymbol \rho}_0[{\bf n}] - {\bf n}\right).
\end{align}
This directional derivative is given by
\begin{align}
\frac{\partial {\bf f}({\bf n} + \lambda {\bf v}_1)}{\partial \lambda} &= \frac{\partial\left( {\boldsymbol \rho}_0[{\bf n} + \lambda {\bf v}_1] -{\bf n} - \lambda {\bf v}_1\right)}{\partial \lambda}\\
 &= \frac{\partial {\boldsymbol \rho}_0[{\bf n} + \lambda {\bf v}_1]}{\partial \lambda} - {\bf v}_1.
\end{align}
The directional derivative can be calculated with the graph-based canonical quantum perturbation theory in Sec.\ \ref{GraphPRT}, where the normalized preconditioned residual charge vector, ${\bf v}_1$, induces a perturbation in the effective single-particle potential. This induced potential then appears as the first-order perturbation in the effective single-particle Hamiltonian, ${\bf H}_1$. This Hamiltonian perturbation can then be partitioned into subsystems using the data dependency graph. The process can be described schematically as,
\begin{align}
    {\bf v}_1 \rightarrow  {\bf H}_1[{\bf v}_1] = \left\{{\bf h}_1^{(k)}\Big \vert_{\rm c} \right\}_{\rm collect}.
\end{align}
We can then determine the density matrix response from the separate subsystems,
\begin{align}
    &{\bf D}_1^\beta\big\vert_{G_\tau} = \left\{ {\bf d}_{1,\lambda}^{\beta,(k)} \Big \vert_{\rm c}\right\}_{\rm collect} + \mu_1 \left\{ {\bf d}_{1,\mu}^{\beta,(k)} \Big\vert_{\rm c}\right\}_{\rm collect},
\end{align}
with the canonical graph-based quantum perturbation theory, as described in Sec.\ \ref{GraphPRT}. Here $\mu_1$ is the shared response in the chemical potential, which is chosen to keep the perturbed full composite system charge neutral. 
The density matrix response, ${\bf D}_1^\beta\big\vert_{G_\tau}$, then determines the electronic charge density response, or the response in the charge vector due to ${\bf v}_1$. However, this can just as well be performed over the separate subsystems, where 
\begin{equation}\begin{array}{l}
     {\displaystyle \frac{\partial {\boldsymbol \rho}_0^{(k)}[{\bf n} + \lambda {\bf v}_1]}{\partial \lambda} = {\boldsymbol \rho}_0^{(k)} \left({\bf d}_{1, {\rm v}_1}^{\beta,(k)}\right) \Big \vert_{\rm c}} \\
    {\displaystyle ~~~ + \mu_1 {\boldsymbol \rho}_0^{(k)} \left({\bf d}_{1,\mu}^{\beta,(k)} \right)\Big \vert_{\rm c}}.
\end{array}
\end{equation}
This allows for a natural parallelism. Here $ {\boldsymbol \rho}_0^{(k)} \left({\bf d}_{1,{\rm v}_1}^{\beta,(k)}\right) \Big \vert_{\rm c}$ is the core part of the charge response determined by the derivative of the susbsystem density matrix, ${\bf d}_{1,{\rm v}_1}^{\beta,(k)}$, with respect to $\lambda$ in Eq.\ (\ref{ddlambda}) with ${\bf h}_1 = {\bf v}_1$.
The next term, ${\boldsymbol \rho}_0^{(k)} \left({\bf d}_{1,\mu}^{\beta,(k)} \right)\Big \vert_{\rm c}$, is the corresponding charge response determined by the density matrix derivative with respect to the chemical potential as in Eq.\ (\ref{dddmu}), and $\mu_1$ is determined as in Eq.\ (\ref{dmu_1}). From this directional response in the charge induced by the charge vector ${\bf v}_1$, we get 
\begin{align}
    {\bf f}_{{\bf v}_1} = \left\{ {\bf k}_0^{(k)} \left(\frac{\partial {\boldsymbol \rho}_0^{(k)}[{\bf n} + \lambda {\bf v}_1]}{\partial \lambda} -{\bf v}_1^{(k)}\right)\right\}_{\rm collect}.
\end{align}
The next vector ${\bf v}_2$ of then Krylov subspace is the constructed from the components of ${\bf f}_{{\bf v}_1}$  orthonormalized to ${\bf v}_1$, i.e.\
\begin{align}
    {\bf v}_2' & ={\bf f}_{{\bf v}_2} - ({\bf f}_{{\bf v}_2}^T{\bf v}_1){\bf v}_1\\
    {\bf v}_2 & ={\bf v}_2'\big / \|{\bf v}_2'\| = \left\{ {\bf v}_2^{(k)}\right\}.
\end{align}
This process of generating orthogonalized Krylov subspace vectors can be repeated iteratively until a sufficiently accurate approximation is achieved \cite{ANiklasson20}. The general scheme can be described by the following algorithm:
\begin{enumerate}
    \item Chose ${\bf v}_1$ from the residual as in Eqs.\ (\ref{v1_prim})-(\ref{v1})
    \item ${\bf v}_i \rightarrow {\bf H}_i[{\bf v}_i] = \left\{{\bf h}_i^{(k)} \Big \vert_{\rm c}\right\}_{\rm collect}$
    \item ${\bf f}_{{\bf v}_i} = \left\{ {\bf k}_0^{(k)} \left(\frac{\partial {\boldsymbol \rho}_0^{(k)}[{\bf n} + \lambda {\bf v}_i]}{\partial \lambda} -{\bf v}_i^{(k)}\right)\right\}_{\rm collect}$
    \item  ${\bf v}_{i+1}' = {\bf f}_{{\bf v}_i} - \sum_{j = 1}^{i-1} ({\bf f}_{{\bf v}_i}^T{\bf v}_j){\bf v}_j$\\
    ${\bf v}_{i+1}  ={\bf v}_{i+1}'\big / \|{\bf v}_{i+1}'\| = \left\{ {\bf v}_{i+1}^{(k)}\right\}_{\rm collect}$
    \item Repeat 2-4 until sufficient convergence is achieved in Eq.\ (\ref{LowRankEEOM}).
\end{enumerate}
Once a sufficiently accurate preconditioned subspace approximation is achieved \cite{ANiklasson20}, the overlap matrix and inverse overlap matrix,
\begin{align}
    O_{ij} &= {\bf f}_{{\bf v}_i}^T{\bf f}_{{\bf v}_j}\\
    {\bf M} &= {\bf O}^{-1},
\end{align}
are formed. The inner products, ${\bf f}_{{\bf v}_i}^T{\bf f}_{{\bf v}_j}$, can be collected in parallel from the separate inner products, ${{\bf f}_{{\bf v}_i}^{(k)}}^T {\bf f}^{(k)}_{{\bf v}_j}$, computed from each core part of the subgraphs.

The integration of the electronic equations of motion can then be performed using the preconditioned Krylov subspace approximation as in Eq.\ (\ref{LowRankEEOM}). This approximation can be performed in parallel on the separate subsystems, where
\begin{align} 
{\bf \ddot n}^{(k)} &\approx - \omega^2 \left(\sum_{ij}^{m < N} {\bf v}_i^{(k)} M_{ij} {{\bf f}_{{\bf v}_j}^{(k)}}^T {\bf k}_0^{(k)}\left({\boldsymbol \rho}_{\rm 0}^{(k)}[{\bf n}]\big \vert_{\rm c} - {\bf n}^{(k)}\big \vert_{\rm c}  \right)\right) \label{ElEOM}
\end{align}
thanks to our graph-based approach. In this way the core parts of the charge acceleration, ${\bf \ddot n}^{(k)}$, can be determined and integrated in parallel for the separate subsystems. 
Equation (\ref{ElEOM}) and how it can be used to integrate the electronic equations of motion for our graph-based shadow potential formulation of extended Lagrangian shadow Born-Oppenheimer molecular dynamics scheme is one of the key results of this paper.

\subsection{Self-consistency acceleration}

The techniques used to approximate the preconditioned kernel in Eq.\ (\ref{ElEOM}) can also be used to accelerate the convergence of the iterative solution of the  electronic ground state. We can do this with the Newton scheme,
\begin{align}
    {\bf n}_{\rm new} =& {\bf n}_{\rm old} - {\bf K} \left({\boldsymbol \rho}_{\rm 0}[{\bf n}_{\rm old}]- {\bf n}_{\rm old} \right),\\
    =& {\bf n}_{\rm old} - \left({\bf K}_0{\bf J}\right)^{-1} {\bf K}_0 \left({\boldsymbol \rho}_{\rm 0}[{\bf n}_{\rm old}]- {\bf n}_{\rm old} \right)
    \label{Newton_SCF}
\end{align}
where the kernel ${\bf K}$ or the preconditioned kernel $\left({\bf K}_0{\bf J}\right)^{-1}$ can be approximated using the preconditioned Krylov subspace approximation as in Eq.\ (\ref{ElEOM}). This allows acceleration of achieving self-consistency over the separate subsystems, where
\begin{equation}\label{SCF_ACC} \begin{array}{ll}
    {\bf n}_{\rm new}^{(k)} = & {\bf n}_{\rm old}^{k)}  - \sum_{ij}^{m < N} {\bf v}_i^{(k)} M_{ij} {{\bf f}_{{\bf v}_j}^{(k)}}^T {\bf k}_0^{(k)} \times  \\
    ~~\\
    & \times \left({\boldsymbol \rho}_{\rm 0}^{(k)}[{\bf n}_{\rm old}]\big \vert_{\rm c} - {\bf n}^{(k)}_{\rm old}\big \vert_{\rm c}  \right). 
    \end{array}
\end{equation}
With an initial guess that is sufficiently close to the ground state in combination with a good preconditioner (kept constant during the SCF iterations) and a sufficient number of low-rank updates, we can expect close to a quadratic convergence. The method is then numerically equivalent to a direct Newton optimization scheme. In general cases, the computational efficiency may not be as good as related Broyden or Pulay mixing schemes \cite{CGBroyden65,DGAnderson65,PPulay80}, but the Newton method in combination with the preconditioned Krylov subspace approximation offers and alternative that may help to accelerate particularly difficult cases. With the graph-based techniques presented in this paper, this Newton-based SCF-acceleration method is now well-suited also for large-scale simulations of complex chemical systems. 

\section{Examples and analysis}

In this section we will show some simulation examples of graph-based quantum response theory and extended Lagrangian shadow Born-Oppenheimer molecular dynamics using the preconditioned Krylov subspace approximation for the integration of the extended electronic degrees of freedom.  First we briefly present SCC-DFTB theory which forms the basis for our implementation. We then demonstrate the graph-based quantum response theory as it is applied to the preconditioned Krylov subspace approximation in the quasi-Newton SCF acceleration scheme in Eq.\ (\ref{SCF_ACC}). Thereafter we demonstrate the methodology for graph-based shadow Born-Oppenheimer molecular dynamics simulations and how we can achieve stability and control of the residual charge errors without performing any SCF optimization prior to the force evaluation as in regular Born-Oppenheimer molecular dynamics.  We then demonstrate how the graph-based methodology can be applied to molecular dynamics simulations of large complex molecular systems with tens-of-thousands of atoms. In all systems we used periodic boundary conditions. At the end we make a preliminary assessment of the scalability and the parallel efficiency. 

\subsection{Implementation with SCC-DFTB theory}

The methods in this article were implemented and tested using semi-empirical SCC-DFTB theory
\cite{DPorezag95,GSeifert96,MElstner98,MFinnis98,TFrauenheim00,MGaus11,BAradi07,BHourahine20} 
based on the open-source electronic structure software package LATTE \cite{LATTE,AKrishnapriyan17} together with the PROGRESS and BML libraries \cite{2016progress,BML}. No new parameterizations or optimizations of the SCC-DFTB energy functional were performed for this study. SCC-DFTB is an approximation of first-principles Kohn-Sham density functional theory, which is derived from a second or third-order expansion of the Kohn-Sham energy functional in the electronic charge fluctuations around a set of overlapping atomic electron densities. A minimal numerical basis set is used and bond and overlap integrals are tabulated and parameterized using a Slater-Koster approximation. The electrostatic energy is approximated by screened Coulomb interactions between atomic net Mulliken partial charges. The Coulomb interaction between atoms decays as $\vert {\bf R}_I - {\bf R}_J\vert^{-1}$ at large distances and is screened at short distances as the interaction between two penetrating Slater-like charge densities with the on-site term chosen as the chemical hardness or a Hubbard U parameter.
In this formalism, the charge density, ${\bf n}$, and the corresponding ground state density, ${\boldsymbol \rho}_0[{\bf n}]$, are reduced to
vectors with components corresponding to the net Mulliken partial electron occupation for each atom.

\subsection{Graph-based accelerated SCF optimization}

The preconditioned Krylov subspace expansion for the kernel can be used to accelerate the convergence in the iterative optimization of the self-consistent ground state solution, Eq.\ (\ref{SCF_ACC}). This ground state SCF optimization is required in our graph-based shadow Born-Oppenheimer molecular dynamics simulations, but only in the first initial time step, where ${\bf n}(t_0)$ is set to the exact regular Born-Oppenheimer ground state, ${\boldsymbol \rho}_{\rm min}({\bf r})$. The SCF acceleration in Eq.\ (\ref{SCF_ACC}) can be used to demonstrate our graph-based quantum response theory for fractional occupation numbers, which is required in the construction of the preconditioner and for the low-rank Krylov subspace approximation of the kernel, ${\bf K}$. A significant reduction in the number of iterations required to reach a high level of convergence would demonstrate the expected performance of the theory. The Krylov subspace expansion we will use is truncated and therefore the kernel is not exact. The preconditioned Krylov subspace expansion is therefore only approximate and in this case the iterative updates using Eq.\ (\ref{Newton_SCF}) correspond to a quasi-Newton scheme.

\begin{figure}[ht]
  \includegraphics[width=0.47\textwidth]{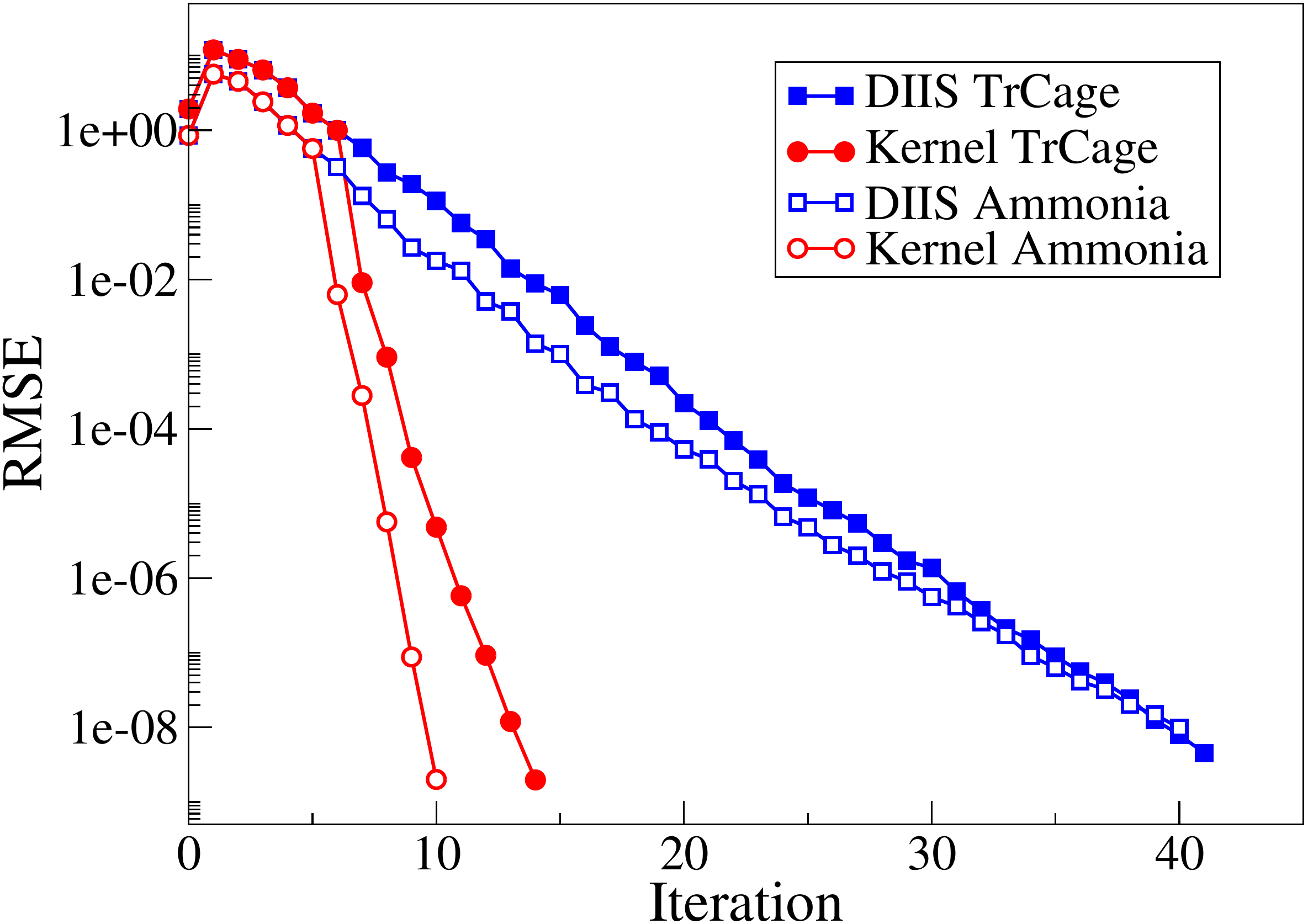}
  \caption{Root mean square error (RMSE) for the partial charges during the iterative self-consistent charge optimization in SCC-DFTB calculations of a solvated Trp-cage protein, as illustrated in Fig.\ \ref{Tr_Cage}, and for ammonia in water, as illustrated in Fig.\ \ref{ammhy}. We are using either Pulay's DIIS mixing scheme \cite{PPulay79,PPulay80} or the preconditioned Krylov subspace approximation for the quasi-Newton scheme (Kernel) in Eq.\ (\ref{Newton_SCF}). The quasi-Newton scheme was initiated after 5 or 6 DIIS iterations.} 
  \label{SCF_acceleration}
\end{figure}

Figure \ref{SCF_acceleration} shows an example of the convergence of the root mean square error (RMSE) given from the root mean square of the residual charge function as a function of iterations, using either the state-of-the-art Pulay direct inversion of the iterative subspace (DIIS) method \cite{PPulay79,PPulay80} or a quasi-Newton scheme (Kernel) as in Eq.\ (\ref{Newton_SCF}). The test systems for this comparison are a solvated Trp-cage protein structure (2,644 atoms) and a system with ammonium hydroxide in water (4,071 atoms). A high electronic temperature was chosen with $\beta^{-1} = 0.5$ eV, which provides a notable deviation in the occupation numbers from a pure ensemble. Both systems are fairly demanding to converge. The quasi-Newton scheme provides a significant acceleration of the SCF optimization and thus demonstrates how the graph-based canonical quantum-response theory works in practice. However, each iteration involves low-rank updates that require repeated quantum response calculations on the graph (a total of six was used for these runs), each which cost about half as much as a full self-consistent field iteration using DIIS. The wall-clock time required to reach convergence is therefore in practice about the same, but only if we ignore the additional cost of calculating the preconditioner. The quasi-Newton scheme represents an alternative acceleration method, for example to the SCF mixing schemes by Broyden, Anderson and Pulay \cite{CGBroyden65,DGAnderson65,PPulay80}, and should be particularly competitive if the cost of the preconditioner can be ignored, which is the case for our graph-based QMD simulations where the preconditioner can be reused over hundreds or even thousands of time steps.

The improved SCF acceleration in Fig.\ \ref{SCF_acceleration} demonstrates that our graph-based canonical quantum response theory and preconditioned Krylov subspace approximation works as expected.  Beyond applications to single-point SCF optimization and molecular dynamics simulations, demonstrated below, the same graph-based quantum response theory and preconditioned Krylov subspace approach should also be applicable to the repeated SCF optimization required in geometry optimization, which also could reuse the same preconditioner over multiple atomic configurations.

\subsection{QMD: Test system A}

To assess the performance of the graph-based shadow Born-Oppenheimer molecular dynamics scheme for more challenging simulations we prepared a test system in which reactions are poised to occur. 
The test system consists of concentrated ammonium hydroxide in water, which in a microcanonical ensemble (NVE) simulation will evolve into ammonia by the reaction: 
$\text{NH}_4^+$ + $\text{OH}^-$ $\rightleftarrows$ $\text{NH}_3$ + $\text{H}_2\text{O}$. This system is highly reactive mainly because of two factors: 1) The concentration of the reactants is artificially high (29.5 Molar or 36 times higher than its equilibrium concentration under ambient conditions) and 2) the initial coordinates are out of equilibrium, which leads to a rapid increase in temperature within the first few hundred MD time steps of the NVE simulation.  

\begin{figure}[ht]
  \includegraphics[width=0.5\textwidth]{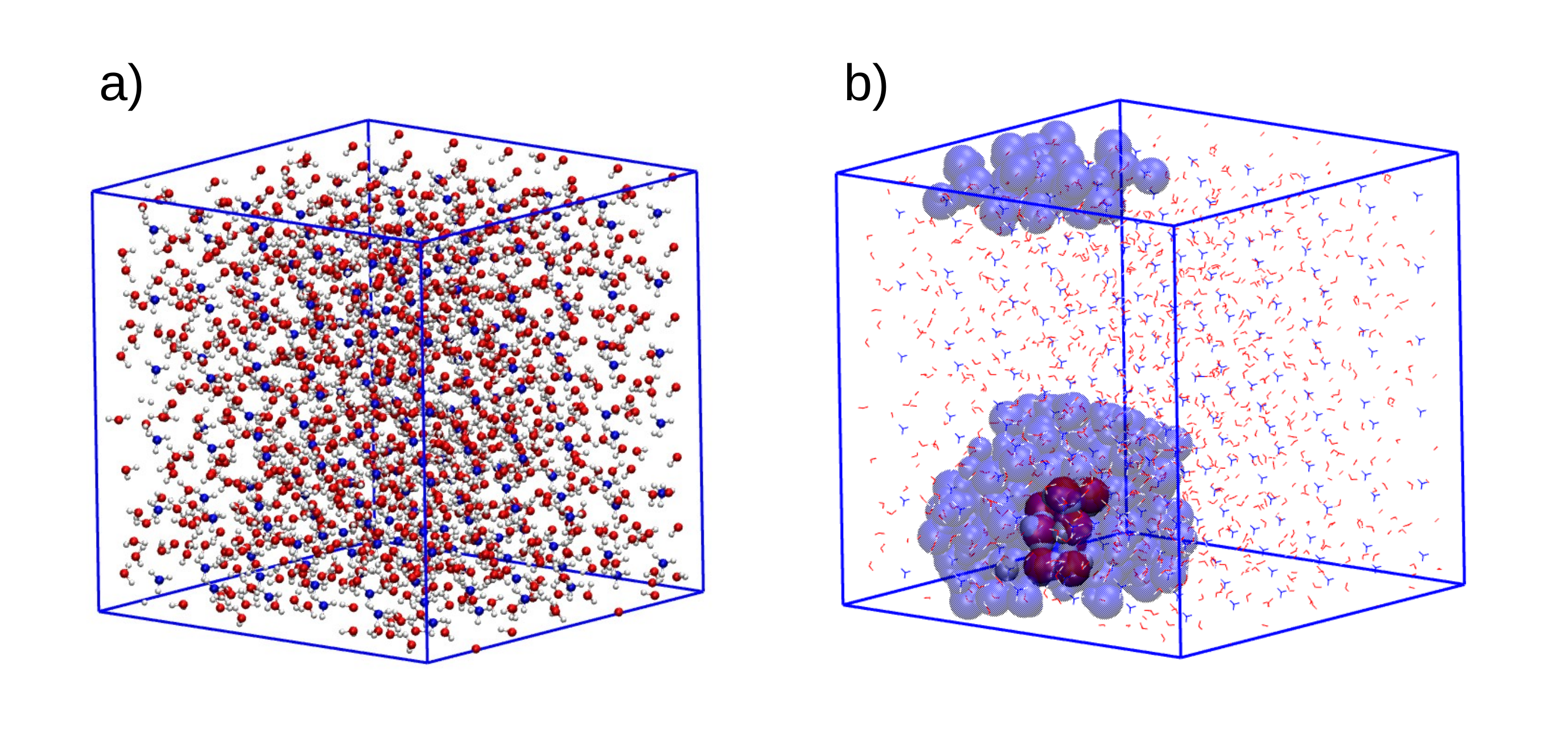}
  \caption{a) Ammonium hydroxide and water in a periodic box containing 4,071 atoms. The system was partitioned using 128 subgraphs. b) A core partition together with its surrounding halo region (in translucent blue large spheres). The halo atoms on the top are near those on the bottom, due to the periodic boundary conditions.}  
  \label{ammhy}
\end{figure}

A schematic representation of the test system is shown in Figure \ref{ammhy} a). It contains a total of 4,071 atoms with an initial set of 216 $\text{NH}_4^+$ + $\text{OH}^-$ pairs apart from the water. 
The data dependency graph was estimated from paths of length two of the combined graph of the full Hamiltonian and a thresholded density matrix (threshold 0.002). We used six rank-1 updates for the Krylov approximation of the preconditioned kernel.
The system was partitioned into 128 subgraphs.  In Figure \ref{ammhy} b) we show one of the subsystems with its core and surrounding halo region, which was automatically selected using a graph-partitioning algorithm as implemented in the METIS software package \cite{metis,metissource}. The QMD simulation was performed using 64 MPI ranks distributed across 16 \texttt{Intel(R) Xeon(R) E5-2695 v4 @ 2.10GHz} CPUs each of them containing 36 cores.

\begin{figure}[ht]
  \label{QMD_Simulation}
  \includegraphics[width=0.55\textwidth]{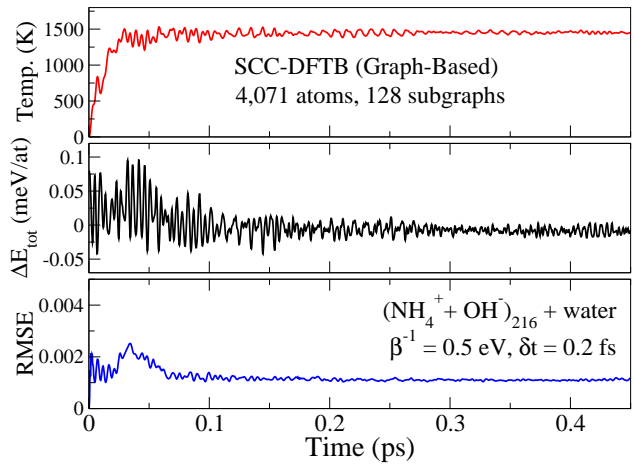}
  \caption{Graph-based QMD simulation of ammonium hydroxide in water containing a total of 4,071 atoms, using extended Lagrangian shadow Born-Oppenheimer molecular dynamics. The initial configuration includes 216 $\text{NH}_4^+$ + $\text{OH}^-$ pairs. The upper panel (red) shows the rapid increase in the statistical temperature. The mid panel (black) shows the oscillations in the total energy (kinetic + potential), and the lower panel (blue) shows the root mean square error (RMSE) given from root mean square of the charge residual function, ${\bf f(n)} = {\boldsymbol \rho}_0[{\bf n}]-{\bf n}$.}  
  \label{ammplots}
\end{figure}

Figure \ref{ammplots} shows three quantities monitored over the QMD simulation. The temperature (in red, upper panel) increases rapidly up to about 1,450 K. The total energy (in black, mid panel) demonstrates a stable total energy without any systematic drift. The root means square error (RMSE) given from the root mean square of the residual charge function (in blue, lower panel) demonstrates a well controlled stable behavior. The rapid initial changes in the electronic structure and temperature creates fairly large oscillations in the total energy, where a few molecules are moving fast. The amplitude of the total energy oscillations decays as the system is reaching a thermal equilibrium. Some of the rapid initial oscillations also lead to a small increase in the residual charge.

\begin{figure}[ht]
  \includegraphics[width=0.5\textwidth]{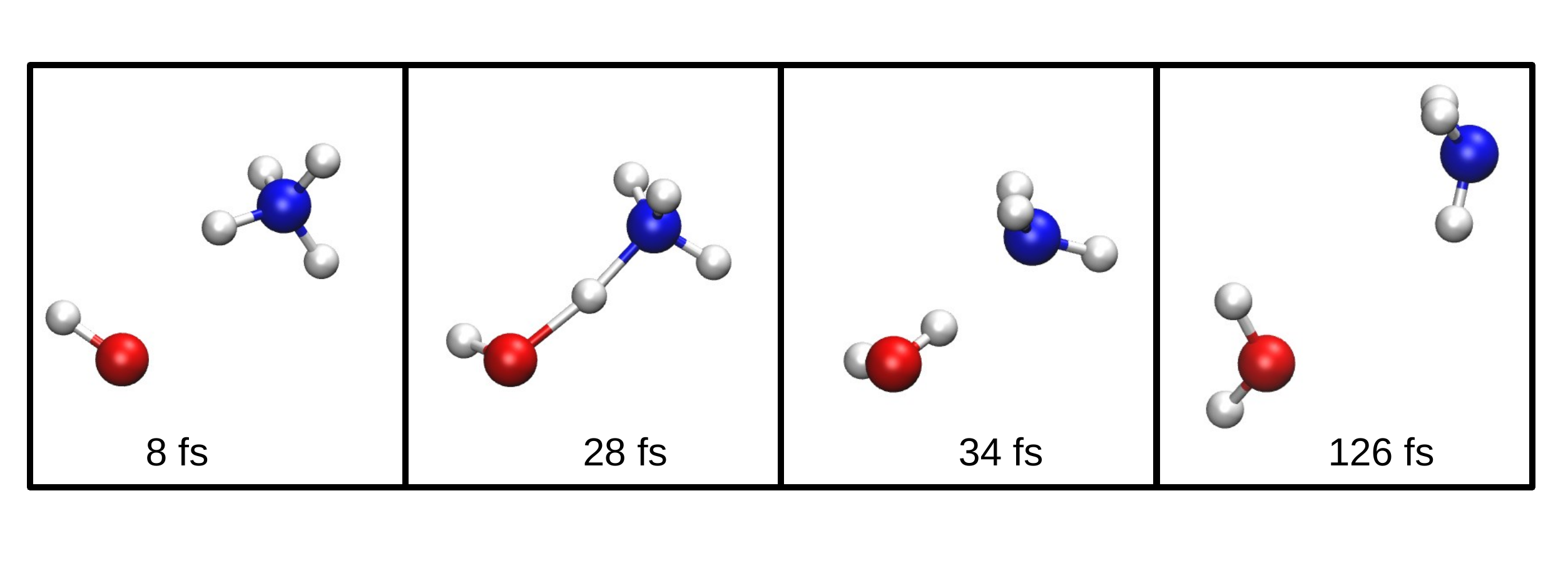}
  \caption{Local molecular snapshots demonstrating the following reaction (from left to right panel): $\text{NH}_4^+$ + $\text{OH}^-$ $\rightarrow$ $\text{NH}_3$ + $\text{H}_2\text{O}$. This was determined from the graph-based QMD simulation in Fig.\ \ref{ammplots}.}  
  \label{ammreact}
\end{figure}

The QMD trajectory was inspected to identify reaction events. In Fig.\ \ref{ammreact} we show a local reaction event captured in a series of configuration snapshots. An initial $\text{NH}_4 - \text{OH}$ complex is formed at 28 fs followed by the formation of $\text{NH}_3$ and water at 34 fs. The rest of the simulation shows $\text{NH}_3$ and water molecules defusing apart (see snapshot at 126 fs).

Despite the out-of-equilibrium initial configuration, followed by a rapid exothermic process with chemical changes as in Fig.\ \ref{ammreact}, the total energy in Fig.\ \ref{ammplots} remains remarkably stable. No instabilities in the total energy or in the residual charge are observed. This example illustrates that the graph-based shadow Born-Oppenheimer molecular-dynamics scheme is capable of simulating a challenging reactive test system without the iterative SCF optimization steps that normally would be required prior the force evaluations in a regular Born-Oppenheimer molecular dynamics simulation.

\subsection{QMD: Test system B}

\begin{figure}[ht]
  \includegraphics[width=0.5\textwidth]{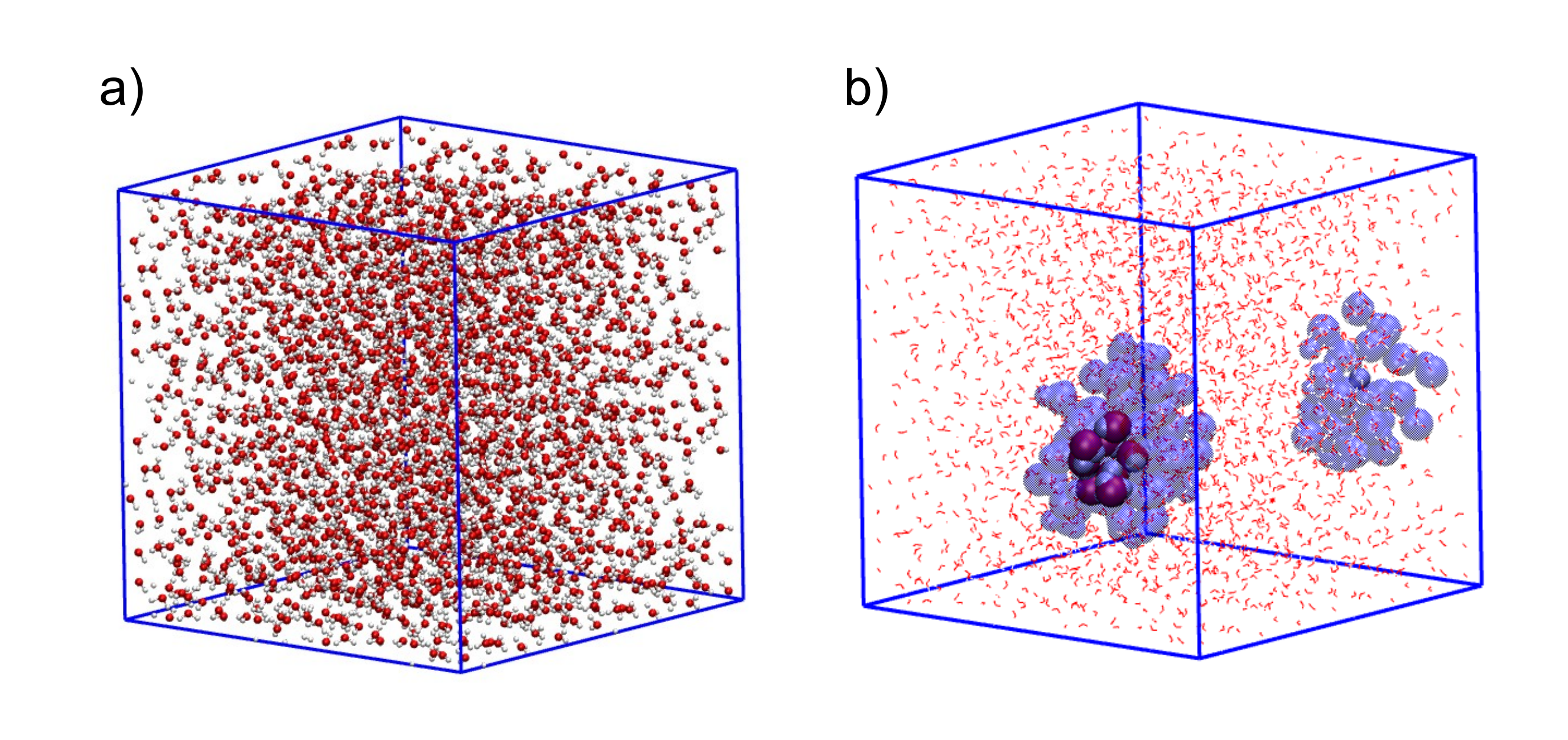}
  \caption{a) Water system in a periodic box containing 6,495 atoms. The system was partitioned using 256 subgraphs. b) A core partitioning (in WDV large sphere representation) together with its halo region (in translucent blue large spheres). The halo atoms near the back-right face are close those near the front-left face, due to the periodic boundary conditions.}  
  \label{wat6K}
\end{figure}

Our next test system was chosen to be a water box containing several thousands of atoms. The system was constructed using the GROMACS solvation tool \cite{GROMACS}. This type of system is non-reactive under the initial conditions and the simulation time-scales (i.e no water dissociation is expected). It therefore allows us to use a relatively coarse graph, leading to more efficient calculations and the possibility of reaching longer-duration simulation times, even for larger systems.  
Figure \ref{wat6K} a) shows the water system containing 6,495  atoms and b) shows a subgraph partition with its core and halo regions. 

\begin{figure}[ht]
  \includegraphics[width=0.53\textwidth]{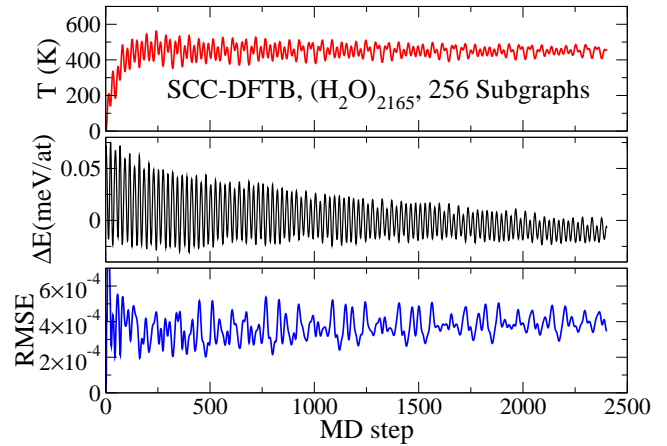}
  \caption{Graph-based QMD simulation of the water system in Fig.\ \ref{wat6K}. The upper panel shows the statistical temperature. The mid panel shows the fluctuations in the total energy per atom, and the lower panel shows the root mean square error (RMSE) calculated from the root mean square of the charge residual function, ${\bf f(n)} = {\boldsymbol \rho}_0[{\bf n}]-{\bf n}$. }  \label{MDwat6K}
\end{figure}

Figure \ref{MDwat6K} shows the result of a NVE simulation starting with an initial out-of-equilibrium structure. The calculations were performed by decomposing the system into 256 subgraphs. 
The QMD simulation was performed using 64 MPI ranks distributed across 16 \texttt{Intel(R) Xeon(R) E5-2695 v4 @ 2.10GHz} CPUs each of them containing 36 cores. MD time step and electronic temperature were set to 0.2 fs and $\beta^{-1} = 0.5$ eV. Also this simulation demonstrates stability and error control in the behavior of the total energy (in black, mid panel) and charge residual error (in blue, lower panel) as the statistical temperature increases to around 400 K (in red, upper panel).

\subsection{QMD: Test system C}

\begin{figure}[ht]
  \includegraphics[width=0.65\textwidth]{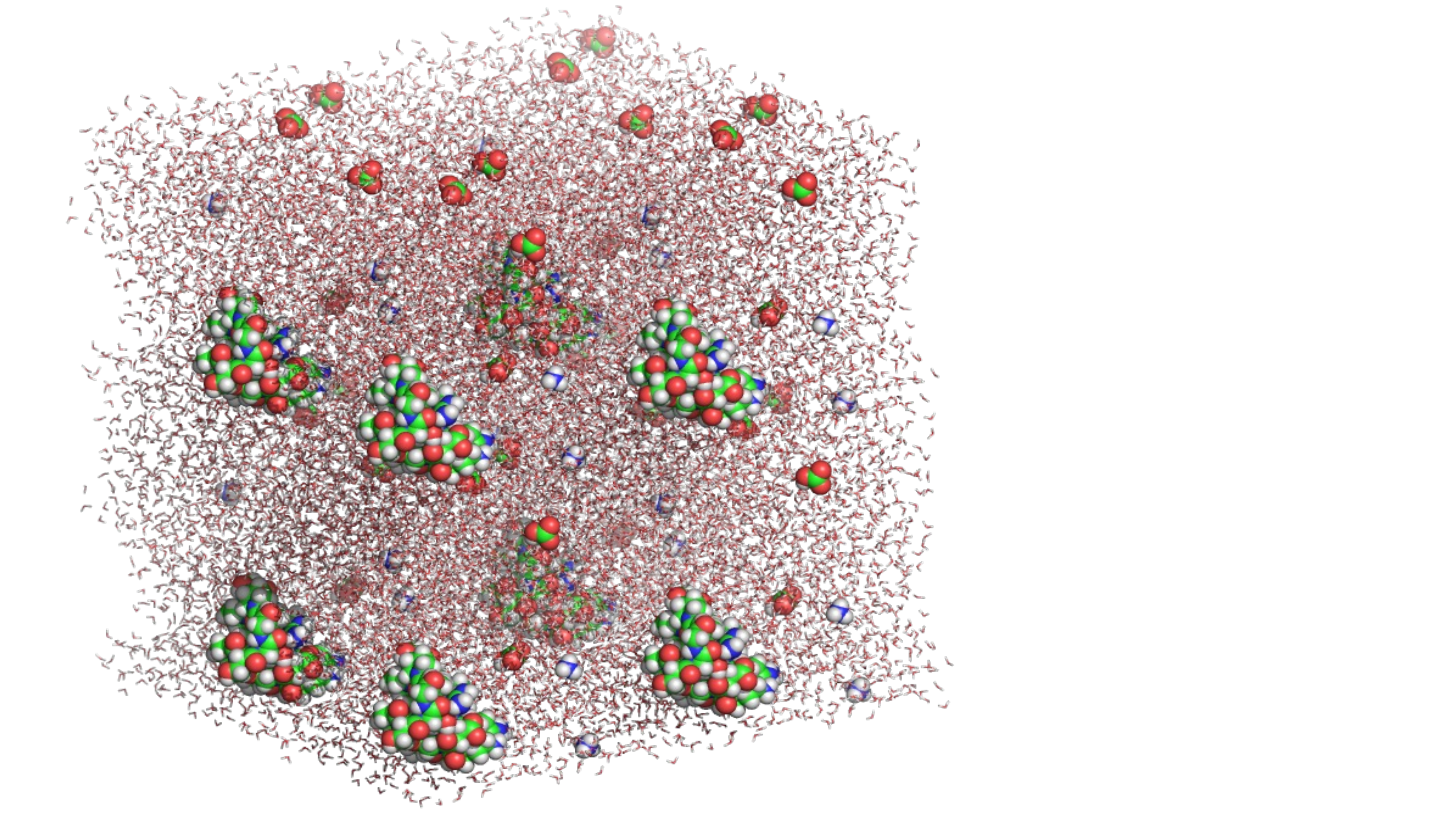}
  \caption{Trp-cage protein molecules with a solution of ammonium bicarbonate and water in a periodic box with a total of 64,112 atoms.}  
  \label{TrpWat64K}
\end{figure}

Our final test case is a system composed by eight Trp-cage synthetic polypeptides in an ammonium bicarbonate solution, shown in Fig. \ref{TrpWat64K}. A solvated simulation box with a single polypeptide and a charge-neutralizing combination of five H$\cdot$CO$_3^{-}$ molecules, and four NH$_4^{+}$ molecules was constructed using the Multicomponent input generator of CHARMM-GUI \cite{charmm_gui}, using input files for the polypeptide, H$\cdot$CO$_3^{-}$, and NH$_4^{+}$. The polypeptide coordinates were obtained from the first model in RCSB PDB entry 1L2Y 
\cite{1L2Y}. 
It was expanded to a 2x2x2 system using PDBProp in AmberTools \cite{AmberTools22}. The final system was a 43.923 $\mathring{A}^3$ cube simulation box with 64,112 atoms. It is highly challenging not to say practically unfeasible to perform quantum-mechanical Born-Oppenheimer simulations of systems of this complexity without super computing access. 

\begin{figure}[ht]
  \includegraphics[width=0.47\textwidth]{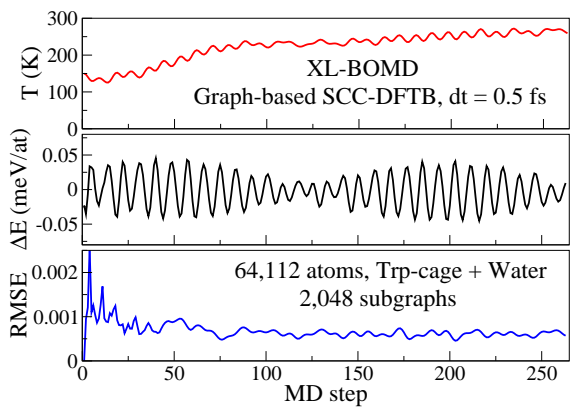}
  \caption{QMD simulation with the graph-based extended Lagrangian shadow Born-Oppenheimer molecular dynamics method of the 64,112 atoms solvated Trp-cage molecules illustrated in Fig.\ \ref{TrpWat64K}. The upper panel shows the statistical temperature, the mid panel the fluctuations in the total energy per atom, and the lower panel shows the root mean square error (RMSE) from the root mean square of the charge residual function, ${\bf f(n)} = {\boldsymbol \rho}_0[{\bf n}]-{\bf n}$. }
  \label{MDTrpWat64K}
\end{figure}

Figure \ref{MDTrpWat64K} shows the results of a QMD simulation with the graph-based extended Lagrangian shadow Born-Oppenheimer molecular dynamics method. The simulation was performed on 32 nodes of the Chicoma Institutional Computing cluster at Los Alamos National Laboratory. Each node is equipped with two 64-core AMD Rome EPYC 7H12 processors and 512 GB memory, with MPI communication via a 100Gb/s HPE/Cray Slingshot10 interconnect (no GPUs). The job was distributed over 1024 MPI tasks: 32 tasks per node, each using 4 OpenMP threads. The computation was distributed using 2,048 subgraphs: 2 per MPI rank. The simulation was performed in a NVE ensemble using a time step of $\delta t = 0.5$ fs, an electronic temperature with $\beta^{-1} = 0.1$ eV, and 3 rank-1 updates in the approximation of the preconditioned kernel. The initial velocity distribution was randomly sampled from a Maxwell distribution with a temperature of 150 K. The size of the core partitions was on average about 30 atoms and the total size of the subgraphs, including the halo, varied between 400 and 500 atoms. Even for this system -- our largest -- the behavior is robust and stable, as seen in the total energy fluctuations (in black, mid panel) and in the size of the charge residual (in blue, lower panel) in Fig.\ \ref{MDTrpWat64K}.  This example demonstrates how the graph-based shadow Born-Oppenheimer molecular dynamics schemes is scalable to system sizes including tens-of-thousands of atoms using a CPU-based compute cluster.

\vspace{1.5cm}
\begin{figure}
  \includegraphics[width=0.45\textwidth]{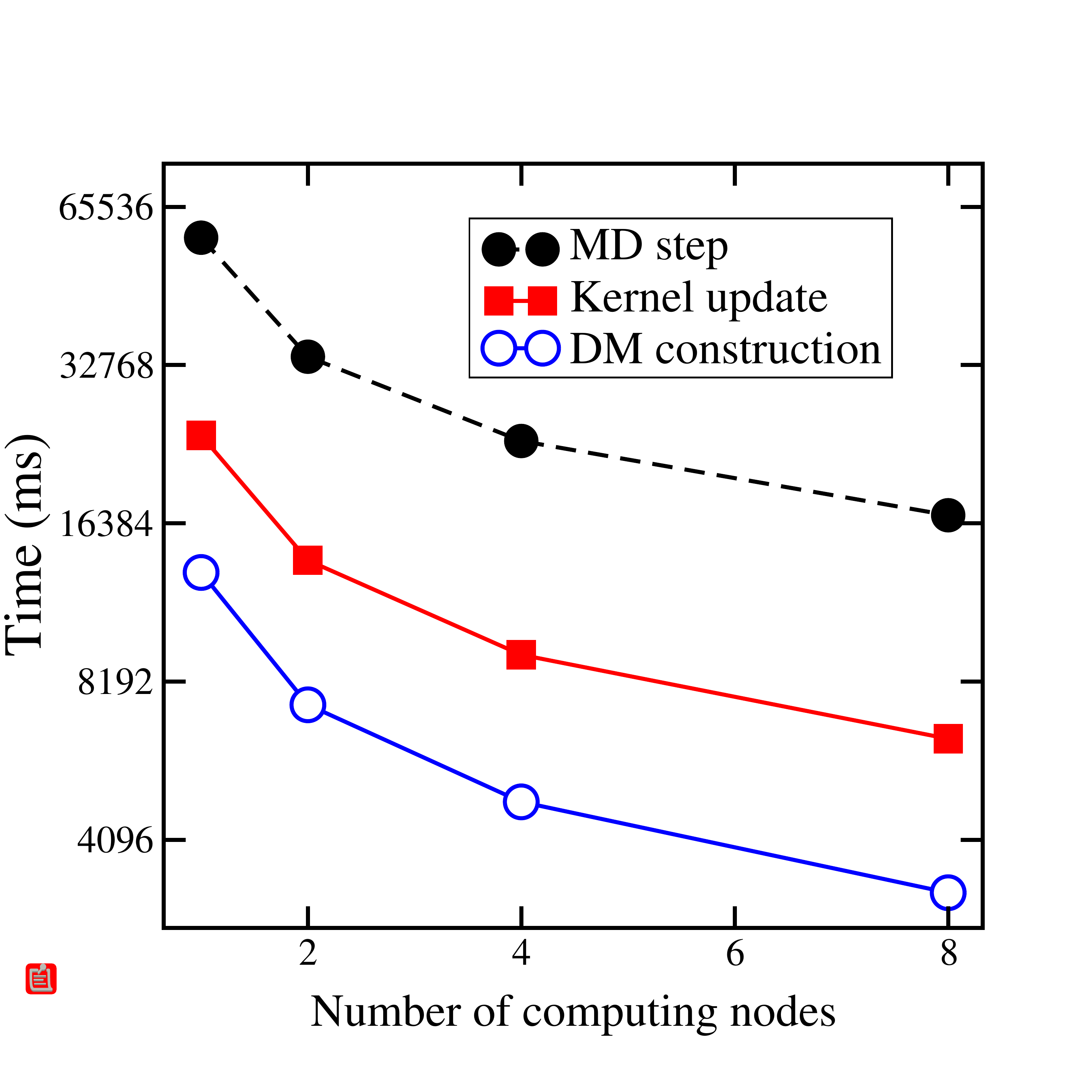}
  \caption{Strong scaling for a single MD step. Time in  milliseconds as a function of the number of computing nodes are shown for the full MD step, the Kernel update, and the density matrix (DM) construction in black, red, and blue respectively. }  
  \label{strongscaling}
\end{figure}

\subsection{Scalability}

To assess the preliminary parallel efficiency of the graph-based approach to electronic structure calculations, we performed both strong and weak scaling studies. The strong scaling study was performed using a 16,704 atom water box, generated by expanding a 2,088 atom box two-fold along each side. The system was subdivided into 512 subgraph partitions. The calculation was distributed on either 1, 2, 4, or 8 nodes of the Chicoma cluster (architecture described in Test System C) using 32 MPI tasks per node, and timings were obtained for the total MD step (including steps that have not yet been prallelized), the kernel update, and the density matrix (DM) construction (Fig.~\ref{strongscaling}). The number of rank updates for the preconditioned Krylov subspace kernel approximation was set to three and the timings were measured after the initial time step, which includes a full SCF optimization of the electronic ground state and the construction of the preconditioner. Compared to using a single node, the total time required for an MD step decreased 1.7-fold, 2.4-fold, and 3.4-fold when increasing the number of nodes to 2, 4, and 8, respectively, corresponding to efficiencies of 85\%, 60\%, and 43\%. Similar strong scaling efficiencies are seen for the kernel update and DM construction steps. 

For the weak scaling study, water boxes were generated as before expanding an initial 2088 water box system. Increasing sizes were prepared and simulated using proportional compute resources. The calculations were performed on a set of homogeneous nodes in the Darwin research testbed cluster at Los Alamos National Laboratory. Each node is equipped with dual socket Intel(R) Xeon(R) CPU E5-2695 v4 $@$ 2.10GHz processors and 125 GB memory. The nodes used are interconnected using 100 Gb/sec Mellanox EDR InfiniBand. The initial 2088 atoms system was used as a reference; this system was simulated using a single compute node. The system was scaled up by factors of 2, 4, 8, and 16, while proportionally increasing the number of nodes. The reference system was decomposed into 64 partitions and the number of partitions was increased in proportion to the system size. This increase in the partitioning may not be the ideal choice for performance, but it simplifies our test. The simulations were performed using four ranks per node. 

\begin{figure}[H]
  \includegraphics[width=0.5\textwidth]{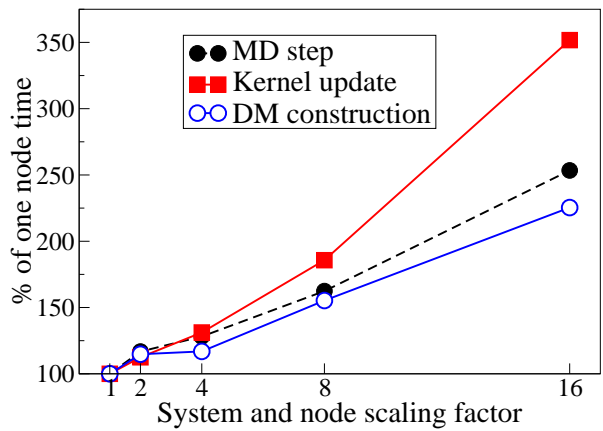}
  \caption{Weak scaling of a single MD step (in black). The relative timings are shown for the Kernel update, the density matrix (DM) construction, and the full molecular dynamics (MD) step.}  
  \label{weakscaling}
\end{figure}

Figure \ref{weakscaling} shows the results of the weak scaling study. Compared to the reference system on a single node, the relative time in the total cost of an MD step increases about 2.5 times for a system 16 times larger using 16 nodes. The density matrix (DM) construction step shows a similar scaling behavior. The Kernel update is comparatively more costly for larger systems: 3.5 times slower when scaling up by a factor of 16.  This reflects an increased level of communication required for the Kernel calculation compared to the other calculations. 

\section{Summary and Conclusions}

In this article we have presented a graph-based canonical quantum perturbation theory for applications in graph-based QMD simulations \cite{ANiklasson16} based on the most recent shadow potential formulations of XL-BOMD \cite{ANiklasson20}. This set of techniques enables stable, linear-scaling, shadow Born-Oppenheimer molecular dynamics simulations of charge sensitive or reactive chemical systems, without involving any iterative SCF optimization prior to the force evaluations.
The formulation includes a preconditioned Krylov subspace approximation for the integration of the extended electronic degrees of freedom, which requires quantum response calculations for the electronic states, including fractional occupation numbers.  The proposed graph-based canonical response calculations can be performed in parallel in the same way 
as for the electronic ground state, which is guided by the partitioning of a data dependency graph that can be estimated from previous integration time steps or SCF iterations.

Graph-based electronic structure calculations are intended for studies of large systems with around a thousand atoms or more and require the separate calculations of subsystems that often contain several hundred atoms. Graph-based QMD simulations are therefore particularly well-suited for semi-empirical electronic structure theory.  
The methods in this article were implemented and tested using SCC-DFTB theory. 

The proposed graph-based shadow Born-Oppenheimer molecular dynamics scheme was demonstrated in simulations of challenging chemically reactive systems such as ammonium hydroxide in water. The simulations were stable both in the fluctuations of the total energy and in the residual charge error. This is in agreement with recent shadow potential formulations of XL-BOMD simulations performed of reactive systems with sensitive unsteady charge solutions, but without using the graph-partitioning scheme \cite{ANiklasson20,ANiklasson20b}. 
We also demonstrated how graph-based canonical quantum response theory in combination with the preconditioned Krylov subspace approximation can be used in a quasi-Newton scheme to accelerate the SCF convergence. 

The graph-based approach to electronic structure calculations can take advantage of emerging exascale computing resources \cite{RSchade22X}. This should be possible also with the graph-based quantum response calculations. Our current implementation used in the examples is still preliminary, but demonstrates good parallel scaling allowing QMD simulations of tens-of-thousands of atoms. More development is needed, in particular, to take advantage of hybrid architectures using graphics processing units or AI-accelerators. The main goal of this paper is the underlying theoretical concepts and techniques for the graph-based electronic structure and response calculations, including fractional occupation numbers, which enable scalable extended Lagrangian shadow Born-Oppenheimer molecular dynamics simulations that also are applicable to simulations of more challenging, charge sensitive or reactive chemical systems. 

Recently it has been shown how statistical machine learning techniques can be used to significantly enhance the accuracy of semi-empirical electronic structure methods \cite{PDral15,Anatole15,DYaron18,JKranz18,NGoldman18,PZheng21,ZGuoqing22,DYaron22}.
To use such ``AI-boosted'' semi-empirical methods in combination with graph-based QMD simulation techniques represents a promising path toward accurate simulations of large complex chemical systems.




\section{Acknowledgements}

This work is supported by the U.S. Department of Energy Office of Basic Energy Sciences (FWP LANLE8AN ``{\em Next Generation Quantum-Based Molecular Dynamics}'')
and by the U.S. Department of Energy through the Los Alamos National Laboratory (LANL), including the LANL Institutional Computing program.
LANL is operated by Triad National Security, LLC, for the National Nuclear Security
Administration of the U.S. Department of Energy under Contract No. 892333218NCA000001.

\bibliography{GraphQMD}

\end{document}



\section*{Supporting information document for: Graph-based Quantum Response Theory and Shadow Molecular Dynamics}

The best way to fully understand the graph-based methodology is probably by following some simple examples. This supporting information illustrates in some detail how a sparse matrix function calculated on any chosen data-dependency graph, G, can be partitioned into functions over separate principal submatrices extracted from the subgraphs of G. A similar demonstration was given in the supporting information of Ref.\ \cite{ANiklasson16}. The calculated results of the separate matrix functions can then be collected either by columns or rows to reproduce the same result as the full sparse matrix function calculated on the graph. This one-to-one correspondence is the key observation behind graph-based electronic structure theory \cite{ANiklasson16}. In these examples, each core partition is only a single vertex corresponding to a diagonal matrix elements. In the first subsection we show how the partitioning is performed for the column case and in the second subsection we show how it can be performed over rows. The examples demonstrates the one-to-one correspondence for non-symmetric matrix polynomials calculated on general directed graphs. LA-UR-22-32368.

\subsection{Matrix polynomial on a graph collected by columns}

This example illustrates the one-to-one relation between a matrixpolynomial, P, calculated on a non-symmetric graph, $G$, and the collected result over submatrix polynomials, $p$, collected by columns.

    \begin{figure}[h]
        \centering
        \includegraphics[width=0.9\textwidth]{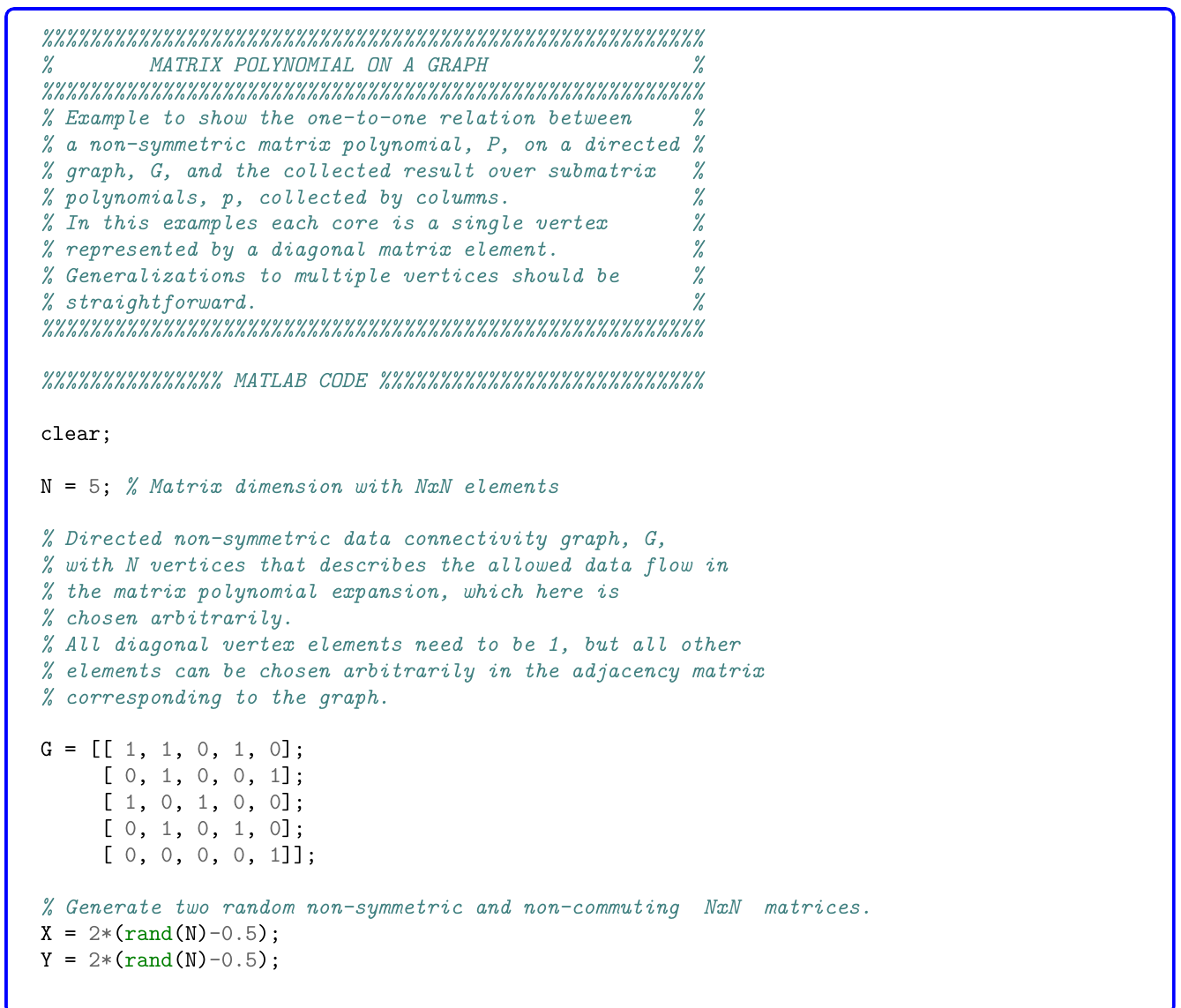}
    \end{figure}
    \begin{figure}
        \centering
        \includegraphics[width=0.9\textwidth]{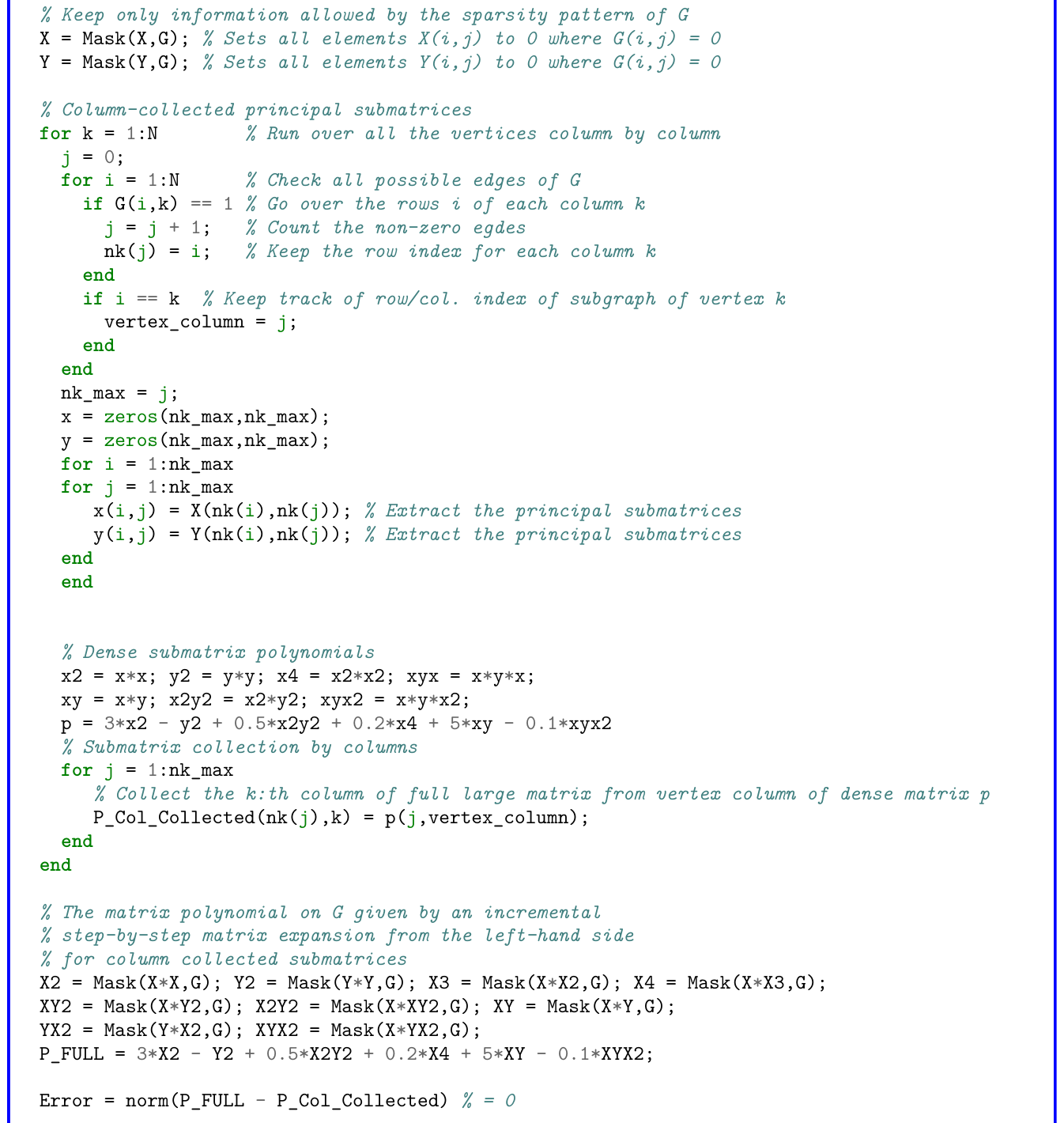}
    \end{figure}
    \begin{figure}
        \centering
        \includegraphics[width=0.9\textwidth]{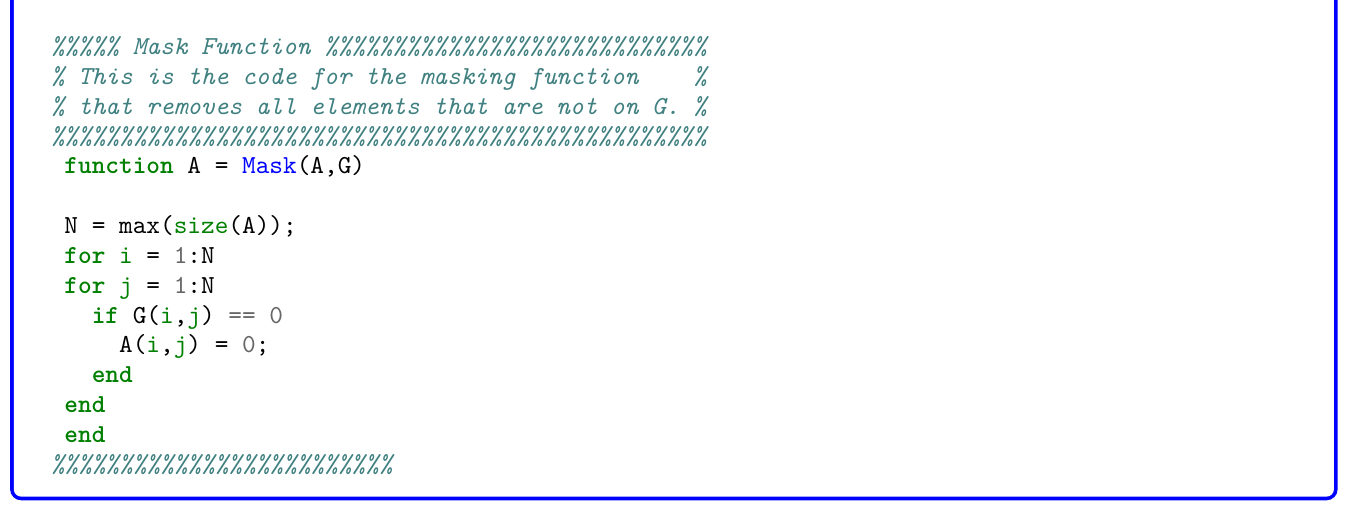}
    \end{figure}

\subsection{Matrix polynomial on a graph collected by rows}

This example illustrates the one-to-one relation between a matrixpolynomial, P, calculated on a non-symmetric graph, $G$, and the collected result over submatrix polynomials, $p$, collected by rows.             %

    \begin{figure}[h]
        \centering
        \includegraphics[width=0.9\textwidth]{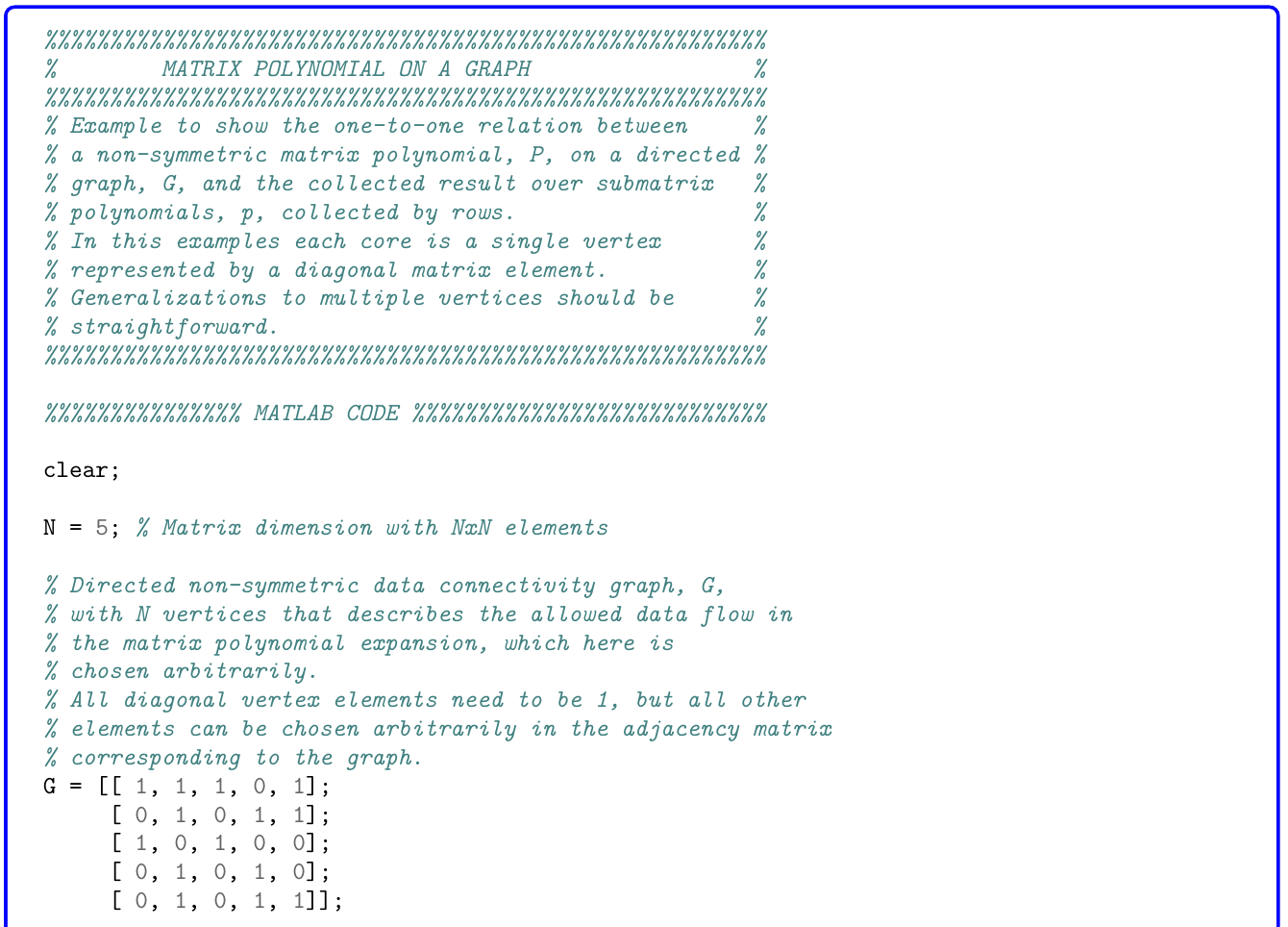}
    \end{figure}
    \begin{figure}[h]
        \centering
        \includegraphics[width=0.9\textwidth]{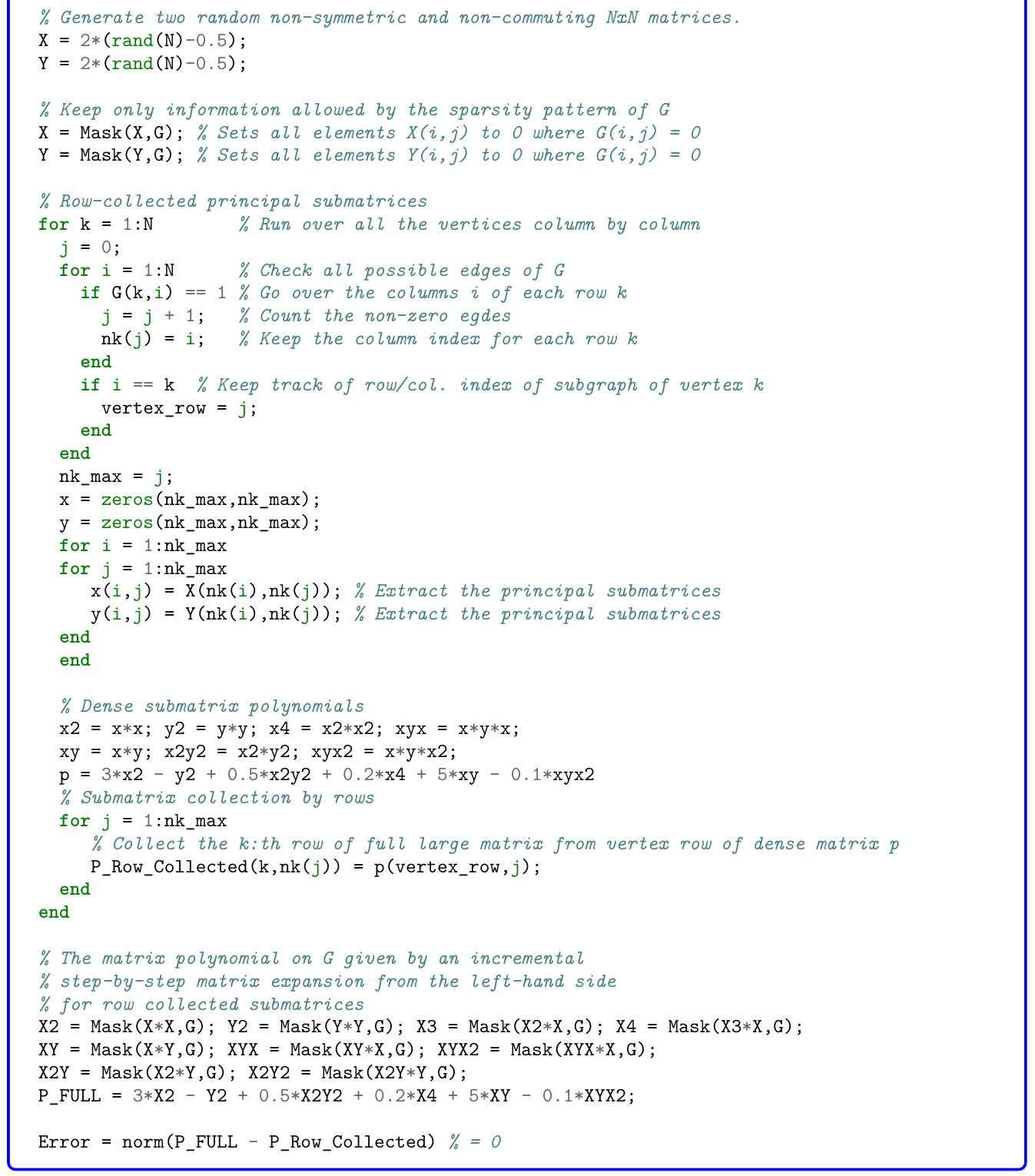}
    \end{figure}

\clearpage
\bibliography{GraphQMD}